\documentclass[aps,prx,twocolumn,showpacs,superscriptaddress,longbibliography]{revtex4-2}
\usepackage[english]{babel}
\usepackage[utf8]{inputenc}
\usepackage{graphicx}
\usepackage{braket}
\usepackage[unicode=true,pdfusetitle,
 bookmarks=true,bookmarksnumbered=false,bookmarksopen=false,
 breaklinks=false,pdfborder={0 0 0},pdfborderstyle={},backref=false,colorlinks=true]
 {hyperref}
 \hypersetup{
 citecolor=blue,
 urlcolor=blue,
 linkcolor=blue}

 \usepackage{latexsym,amsmath,verbatim}
\usepackage{amssymb} 
\usepackage{amsfonts}
\usepackage{colortbl}
\usepackage{array}
\usepackage{stackrel}
\usepackage{bm}
\usepackage{nicefrac}
\usepackage{rotating}
\usepackage{float}
\usepackage[final]{pdfpages}
\usepackage{lipsum}  

\makeatletter
\AtBeginDocument{\let\LS@rot\@undefined}
\makeatother

\begin{document}

\title{Laplacian Renormalization Group for heterogeneous networks}

\author{Pablo Villegas}
\affiliation{'Enrico Fermi' Research Center (CREF), Via Panisperna 89A, 00184 - Rome, Italy}
\author{Tommaso Gili}
\affiliation{IMT Institute for Advanced Studies, Piazza San Ponziano 6, 55100 Lucca, Italy.}
\author{Guido Caldarelli}
\email[Corresponding author: guido.caldarelli@unive.it]{}
\affiliation{Department of Molecular Sciences and Nanosystems, Ca’ Foscari University of Venice, 30172 Venice, Italy}
\affiliation{European Centre for Living Technology, 30124 Venice, Italy}
\affiliation{Institute for Complex Systems, Consiglio Nazionale delle Ricerche, UoS Sapienza, 00185 Rome, Italy}
\affiliation{London Institute for Mathematical Sciences, W1K2XF London, United Kingdom}
\author{Andrea Gabrielli}
\affiliation{'Enrico Fermi' Research Center (CREF), Via Panisperna 89A, 00184 - Rome, Italy}
\affiliation{Dipartimento di Ingegneria, Università Roma Tre, 00146, Rome, Italy}
\affiliation{Institute for Complex Systems, Consiglio Nazionale delle Ricerche, UoS Sapienza, 00185 Rome, Italy}

\vspace{0.5cm}

\vspace{0.5cm}

\renewcommand\refname{} % required to avoid the bibliography heading

\begin{abstract}
The renormalization group is the cornerstone of the modern theory of universality and phase transitions, a powerful tool to scrutinize symmetries and organizational scales in dynamical systems. However, its network counterpart is particularly challenging due to correlations between intertwined scales. To date, the explorations are based on hidden geometries hypotheses. Here, we propose a Laplacian RG diffusion-based picture in complex networks, defining both the Kadanoff supernodes' concept, the momentum space procedure, \emph{\`a la Wilson}, and applying this RG scheme to real networks in a natural and parsimonious way.
\end{abstract}

%Renormalization group is one of the most powerful tools to scrutinize symmetries and organizational scales in dynamical systems, being the cornerstone to develop the modern theory of universality and phase transitions. Nonetheless, the developing of RG techniques in complex structures induced by diffusion distances  has been unexplored up to now, being subjected to difficulties like correlations between intertwined scales and subject to the hypothesis of hidden geometries. Here we proposed a real-space RG picture in complex networks based on information diffusion properties on top of the network by using the Laplacian matrix. We define in a natural and parsimonious way the concept of Kadanoff supernodes, opening the door to the characterization of complex networks in university classes and applying our methods to real well-known scale-invariant networks.´
\maketitle
%%%%%%%%%%%%%%%%%%%%%%%%%%%%%%%%%%%%%%%%%%%%%%%%%%%%%%%%%%%%%%%%

A fundamental open question is how to perform network reduction to generate replicas that connect characteristic internal scales that significantly exceed the microscopic ones. This scenario is the hunting ground for one of the most powerful tools in modern theoretical physics, the Renormalization Group (RG) \cite{Fisher1974, Wilson1974}. RG provides an elegant and precise theory of criticality and allows for connecting --via the scaling hypothesis-- extremely varied spatiotemporal scales and understanding the fundamental issues of scale invariance \cite{Binney, Amit, Kardar}. However, its complex networks counterpart is, to date, slightly studied. Different techniques have been employed, such as spectral coarse-graining \cite{Gfeller}, or box-covering methods \cite{Song2005, Song2006, Goh2006, Kim2007}, which allow identifying general sets of scaling relations in complex networks \cite{Radicchi2008, Rozenfeld2010}, starting with the general assumption of inherent fractal properties of the system. Nevertheless, small-world effects reflected in short path lengths overcomplicate the identification of 'block-nodes' \cite{WS, Klemm2002}, while Kadanoff's decimation present several problems when applied to actual graphs \cite{Garlaschelli}. 

%It allows in understanding how several fundamental units interact at very different spatial, time, and functional scales. %and representing a fundamental step to identify universal properties based only on relevant ingredients: i.e., symmetries, conservation laws, and dimensionalities. 

Solid efforts in the complex network community have been made to develop subsequent RG techniques. In a pioneering work, Garc\'ia-P\'erez et al. \cite{Garcia2018} defined a geometric RG approach by embedding complex networks into underlying hidden metric geometrical spaces. It founds particular application in analyzing RG processes in the Human Connectome network \cite{Zheng2020}, by studying zoomed-out layers and showing that they remain self-similar under particular coarse-graining transformations \cite{Zheng2020, Garcia2018}. Notwithstanding the power of these procedures, they all rely on the critical assumption that they dwell in different isomorphic geometric spaces ($\mathbb{S}^{1}$ and $\mathbb{H}^{2}$), or consider a fitness distance between nodes \cite{Garlaschelli, Radicchi2008}, constantly conditioning the probability of connection among nodes to establish subsequent supernodes. For example, it drives to non-conservation of the average degree along with the RG flow, leading to forced pruning of links in network reduction \cite{Garcia2018}. In particular, developing free-metric RG approaches induced by diffusion distances remains a basic open challenge \cite{Boguna2021}.

Free field or Gaussian theories \cite{Kosterlitz, ZinnJustin}, where the average over randomness generates effective interactions between 'particles' \cite{Gardner}, allowed to make Kadanoff's intuitive ideas quantitative \cite{Kadanoff1966, Wilson1974}. In this specific case, the RG is profoundly linked with diffusion equations \cite{Matsumoto2020}, which, in the particular case of graphs, takes the form of the Laplacian matrix, whose spectral properties determine many dynamical processes on networks \cite{Masuda2017}. The lack of intrinsic diffusion scales, in addition to those of the network structure, made network geometry and topology to be naturally encoded in the spectral properties of the graph Laplacian \cite{Ginestra2020} as, e.g., the spectral dimension of the graph. Recently it has been proven to be essential to identify critical scales \cite{Domenico2016} and core structures of complex networks regardless of their intrinsic properties \cite{Villegas2022}.

Here, we propose a brand-new diffusion-based RG scheme, taking advantage of the Laplacian graph operator, which detects appropriate spatio-temporal scales in heterogeneous networks.
In particular, first, we formulate a heuristic real space version of the RG in which, in analogy with the Migdal-Kadanoff RG prescription \cite{migdal1976,Kadanoff1966}, we define a recursive coarse-graining procedure of the network nodes conserving the diffusion properties of the network at larger and larger spatio-temporal scales. Then, in the spirit of real-space RG techniques \cite{Kadanoff1966}, we introduce the concept of Kadanoff supernodes based on the characteristic resolution scales of the system.
This method overcomes small-world issues and solves decimation problems in performing downscaled replicas. 
We then move to a more rigorous formulation of the diffusion-driven RG, which is analogous to the statistical field theory of the momentum space RG {\em \`a la Wilson}. This consists of formulating a new Laplacian RG (LRG) theoretical framework where fast diffusion modes are progressively integrated out from the Laplacian operator, which automatically induces a definition of coarse-grained macronodes and connections, and finally of a renormalized "slow" Laplacian operator on the coarse-grained graph. Next, we apply the LRG to several real scale-free networks, showing the ability of the method to perform network reduction and capture essential properties of several systems. Finally, in the Methods section, we derive the induced relation between the Laplacian spectrum and the specific heat behavior for scale-invariant networks.
%from very different spheres

%STATISTICAL MECHANICS
\section*{Statistical physics of information network diffusion}
Information communicability in complex networks is governed by the Laplacian matrix  \cite{Newman,Masuda2017}, defined for undirected networks as $L_{ij}=[ (\delta_{ij}\stackrel[k]{}{\sum}A_{ik})-A_{ij}]$, where $A_{ij}$ are the elements of the network's adjacency matrix $A$, and $\delta_{i,j}$ is the Kronecker delta function. The evolution of information of a given initial specific state of the network, $\textbf{s}(0)$, will evolve with time as $\textbf{s}(\tau)=e^{-\tau \hat L}\textbf{s}(0)$. The \emph{network propagator},  $\hat K=e^{-\tau \hat L}$, represents the discrete counterpart of the path-integral formulation of general diffusion processes \cite{Feynman,ZinnJustin}, and each matrix element $\hat K_{ij}$ describe the sum of diffusion trajectories along all possible paths connecting nodes i and j at time $\tau$ \cite{moretti2019}. To fulfill the ergodic hypothesis, we assume the connected networks case.

In terms of the network propagator (see Methods), $\hat K$, it is possible to define the ensemble of accesible information diffusion states \cite{Domenico2016,ghavasieh2020,Villegas2022}, namely,
\begin{equation}
 \mathbf{\rho(\tau)}= \frac{\hat K}{\mathrm{Tr}(\hat K)}=\frac{e^{-\tau \hat L}}{Tr(e^{-\tau \hat L})}\,.
 \label{EvolMat}
\end{equation}
where $\rho(\tau)$ is tantamount to the canonical density operator in statistical physics (or to the functional over fields configurations) \cite{Binney,Pathria,Greiner}. It follows that $S[\rho(\tau)]$ corresponds to the canonical system entropy \cite{Domenico2016,Villegas2022},
\begin{equation}
 S[\rho(\tau)]=-\frac{1}{\log(N)}\stackrel[i=1]{N}{\sum}\mu_{i}(\tau)\log\mu_{i}(\tau).
 \label{S-rho}
\end{equation}

where $\mu_i$ represents the specific $\rho(\tau)$ set of eigenvalues.
In particular, $S\in[0,1]$, reflects the emergence of \emph{entropic transitions} (or information propagation transitions, i.e., diffusion) over the network \cite{Villegas2022}. By increasing the diffusion time $\tau$ from $0$ to $\infty$,  $S[\rho(\tau)]$ decreases from $1$ (\emph{segregated} and heterogeneous phase -- the information diffuses from single nodes only to the local neighborhood) to $0$ (\emph{integrated} and homogeneous phase -- the information has spread all over the network). The temporal derivative of the entropy, $C=-\frac{dS}{d(\log \tau)}$, represents the specific heat of the system, tightly linked with the system correlation lengths. In particular, a constant specific heat is a reflection of the scale-invariant nature of the network (see Methods).

\begin{figure}[hbtp]
 \includegraphics[width=1.0\columnwidth]{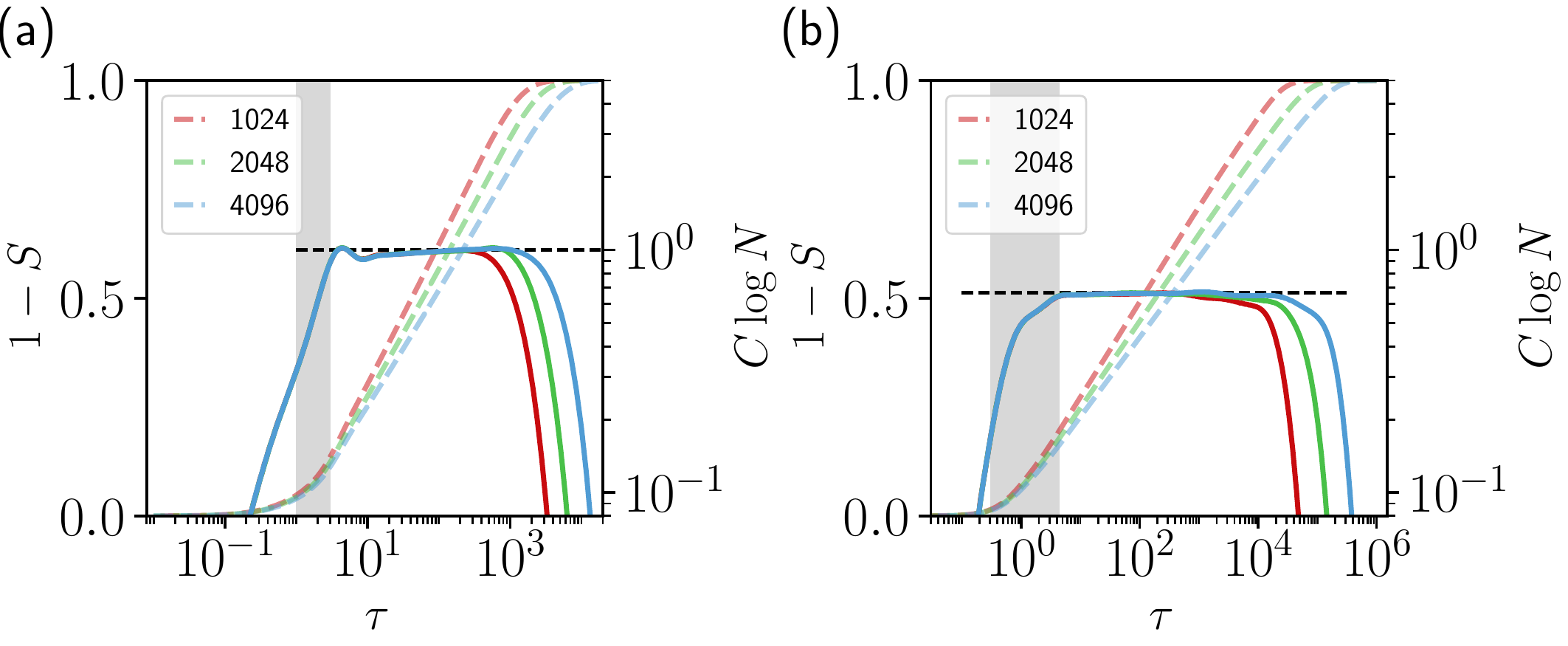}
 \caption{Entropy parameter (dashed lines, $(1-S)$), and specific heat (solid lines, $C$), versus the temporal resolution parameter of the network, $\tau$ for: \textbf{(a)} BA scale-free networks with $m=1$ and, \textbf{(b)} random trees.  Grey area represents the optimal scales to perform Kadanoff supernodes (see below), while black dashed line are the expected analytical specific heat value for both networks (see Methods). Curves have been averaged over $10^2$ realizations.}\label{Entropies}
\end{figure}

As shown in Figure \ref{Entropies}, for the specific case of Barabasi-Albert (BA) networks and random trees (RT), there is a characteristic loss of information as the time $\tau$ increases. The larger the time, the lower the localized information on the different mesoscale network structures. From the analysis of the changes in the entropy evolution (see Methods), together with its derivative, $C$, the characteristic network resolution scales emerge \cite{Villegas2022}. Specifically, the peaks in the specific heat reveal the full network scale at significant diffusion times (scaling with the system size) and the short-range characteristic scales of the network ($\tau^*$, tantamount to the lattice spacing $a$ in well-known Euclidean spaces $\mathbb{R}^n$).

%\textbf{(a)} an Erd\H{o}s-Renyi network of $<k>=30$ and different system sizes (see legend),
%PATH INTEGRALS
\section*{Real-space Laplacian Renormalization Group}
A crucial point is to extract the network 'building blocks', i.e., to generate a \emph{metagraph}, at each time $\tau$ to link the different network mesocales. Note that, at time $\tau=0$, $\rho$ is the diagonal matrix $\rho(0)_{ij} = \delta_{ij} /N$. Hence, $\rho(\tau)$ will be subject to the properties of the network Laplacian, ruling the current information flow between nodes, and reflects the renormalization group flow. So far, we need to consider a rule to scrutinize the network substructures at all resolution scales (i.e., $\tau$). For the sake of simplicity, we choose the following one: two nodes reciprocally process information when they reach a greater than or equal value than the information contained on one of the two nodes  \cite{Villegas2022}, thereby introducing $\rho'=\frac{\rho_{ij}}{\min(\rho_{ii},\rho_{jj})}$. Thus, depending on their particular $\rho_{ij}$ matrix element at time $\tau$, it is possible to define the meta graph, $\zeta_{ij}=\Theta(\rho'-1)$, where $\Theta$ stands for the Heaviside step function. As expected, for $\tau \to \infty$, $\rho$ converges to $\rho_{ij}=1/N$, and $\zeta$ becomes the all-ones matrix.

For a given scale, the meta graph $\zeta$ is thus the binarized counterpart of the canonical density operator, in analogy to the path integral formulation of general diffusion processes \cite{Graham1977}. Note that, after examining all continuous paths traveling along the network \cite{ZinnJustin} and starting from node $i$ at time $\tau=0$, our particular choice selects the most probable paths from  Eq.(\ref{EvolMat}), giving information about the prominent information flow paths of the network in the interval $0<t<\tau$. In the jargon of the statistical mechanics, we are considering the analogy to the Wiener integral and building the RG flow of the network structure \cite{Wilson1974, ZinnJustin}. The last step is how to recursively group the network's nodes into subsequent supernodes, i.e., how to perform decimation. 

%To test the null hypothesis, we check the entropy transitions for regular 1D and 2D lattices, the simplest non-significant scale-invariant spatial structures. Unsurprisingly, lattice structures present a clean plateau across multiple scales when $(\frac{dS}{d(\log\tau}))$ is analyzed (see Appendix \ref{LattAp}), also evidencing a well-behaved spatial-dependent curve collapse.
\begin{figure}[hbtp] 
\begin{center}
 \includegraphics[width=1.0\columnwidth]{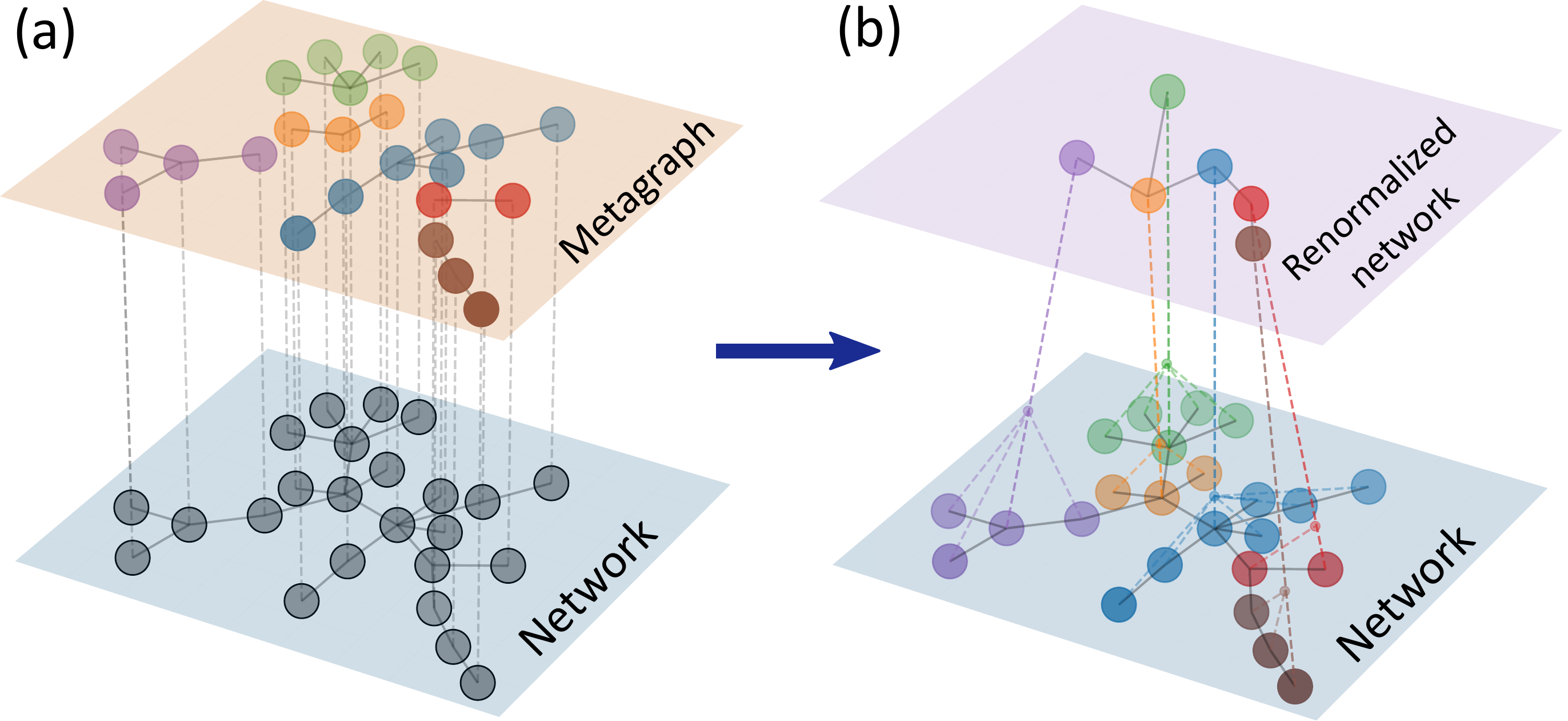}
 \caption{Sketch of the Kadanoff supernodes procedure. (a) Lower layer shows the case of a BA network ($N=24$, $m=1$), and the upper layer, $\zeta$ for $\tau=1.96$. Different colors identify the Kadanoff supernodes. (b) Each block becomes a single node incident to any edge to the original ones.}\label{MetaG}
 \end{center}
\end{figure}
%KADANOFF BLOCKS
In full analogy with the Kadanoff picture, it is possible to consider nodes --under the accurate selection of particular blocking scales of the network-- within regions up to a critical mesoscale, which behaves like a single supernode \cite{Kadanoff1966, Moloney}. Analogously to the real-space RG, there is no unique way to generate new groups of supernodes or coarse-graining, but if the system is scale-invariant, we expect it to be unaffected by RG transformations. In this perspective, using the specific heat, $C$, we propose an RG rule over scales $\tau\sim\tau^*$, where $\tau^*$ stands for the $C$ peak at short-times, realizing the small network scales. The procedure consists of the following steps (see also Fig.\ref{MetaG}):

\begin{enumerate}
    \item Build the network meta graph composed of heterogeneous disjoint blocks of $n_i$ nodes for $\tau\sim\tau^*$, as established by the information network defocusing  \cite{Villegas2022}.
    \item Replace each block of connected nodes with a single supernode.
    \item Consider supernodes as a single node incident to any edge to the original $n_i$ nodes.
    \item Renormalize
\end{enumerate}
% Analogously, it should bring into focus the importance of different observables when performing the RG transformation, i.e., \emph{relevant, irrelevant, and marginal observables}.

\begin{figure*}[hbtp]
 \includegraphics[width=1.9\columnwidth]{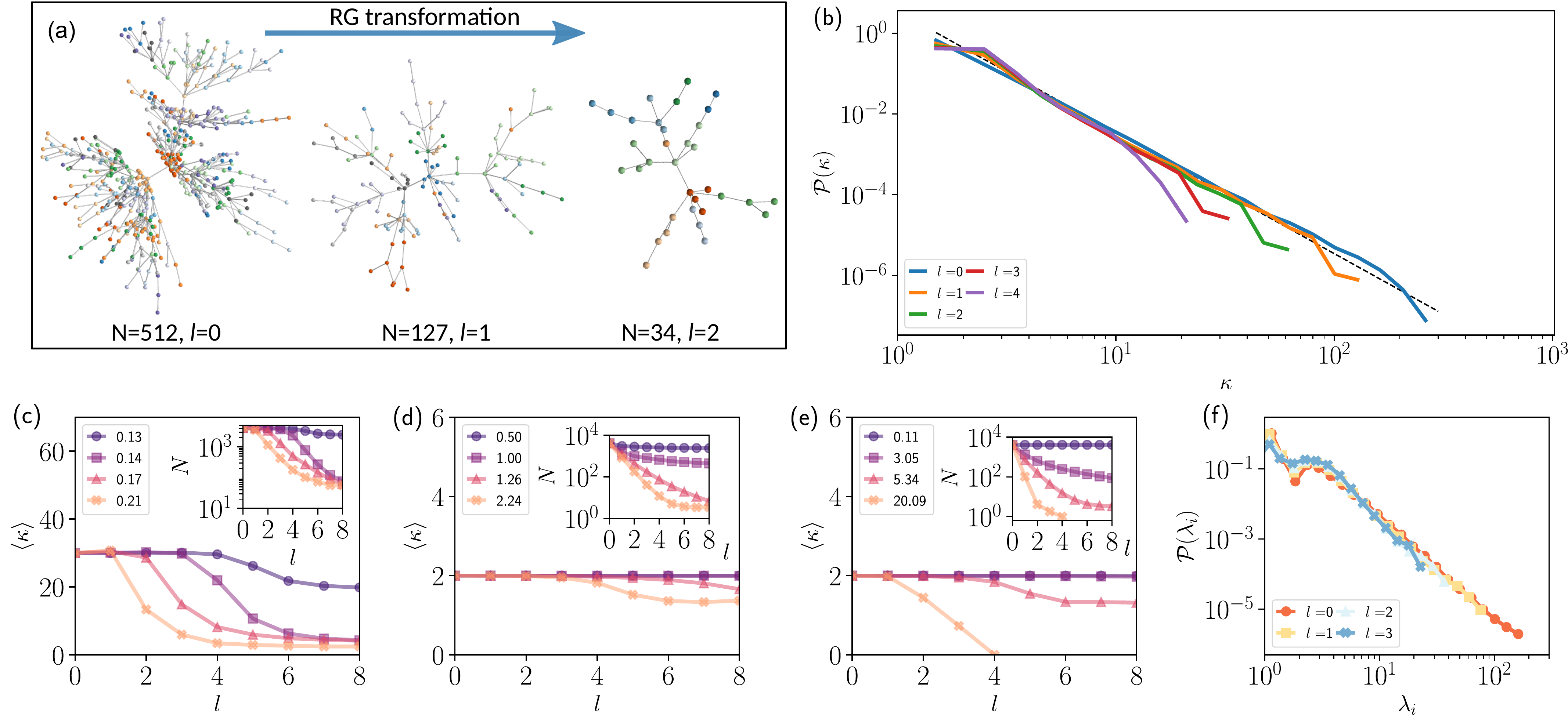}
 \caption{\textbf{Kadanoff supernodes RG. (a)} RG transformation for a particular selection of a BA network ($N=512$, $m=1$). Kadanoff supernodes are plotted in a different color for every scale. \textbf{(b)} Degree distribution for a BA network (solid lines) and a scale-free network with $\gamma=2$ (dashed lines) at different RG steps with $\tau=1.26$ (see legend). \textbf{(c-f)} Mean connectivity flow under subsequent RG transformations for different $\tau$ values (see legend): \textbf{(c)} an Erd\H{o}s-Renyi network of $<k>=30$,  \textbf{(d)} a BA scale-free network with $m=1$ and, \textbf{(e)} random tree. \textbf{(g)} Spectral probability distribution of the downscaled Laplacian replicas, $\hat L_i$, for different RG steps in a BA network (see legend).  All curves have been averaged over $10^2$ network realizations with $N=4096$ at the initial scale.}\label{KBlocks}
\end{figure*}

Figure \ref{KBlocks} shows the RG transformation for many different networks and multiple steps $l$. In the case of Erd\"{o}s-R\'enyi networks, we find that there exists only a characteristic resolution scale (see SI2 \cite{SI}). Kadanoff supernodes select only single nodes at smaller scales, making the network trivially invariant. For any possible grouping of nodes --at every scale-- the mean connectivity of the network decreases after successive RG transformations. The network thus flows to a single-node state, reflecting the existence of a well-defined network scale (see further analysis and other test-cases as, e.g., stochastic block models in SI2 \cite{SI}). 

RG transformation can also be applied to challenging networks of particular interest to real-life applications as small-world ones, revealing the possibility of making network reduction in this type of structures (even if they present yet isolated scales as depicted in SI2 \cite{SI}). Nonetheless, when performing RG analyses over bonafide scale-invariant networks, as the BA model, both the mean connectivity and the degree distribution remain invariant after successive network reductions, conserving analogous properties to the original one (see Fig. \ref{KBlocks} and SI4 \cite{SI} for further analysis). Figure \ref{KBlocks}a shows a graphical example of a three-step decimation procedure for a BA network with $m=1$ and $N=512$ nodes. Different colors at every transformation represent Kadanoff supernodes. Analogously, Fig. \ref{KBlocks}e shows the RG procedure over random trees, confirming the capability of our RG procedure to perform network reduction on top of well-defined synthetic scale-invariant networks (see also  SI3 \cite{SI}). Furthermore, Fig. \ref{KBlocks}f displays the scale-invariant nature of the Laplacian for different downscaled BA replicas.

\begin{figure}[hbtp] 
\begin{center}
 \includegraphics[width=1.0\columnwidth]{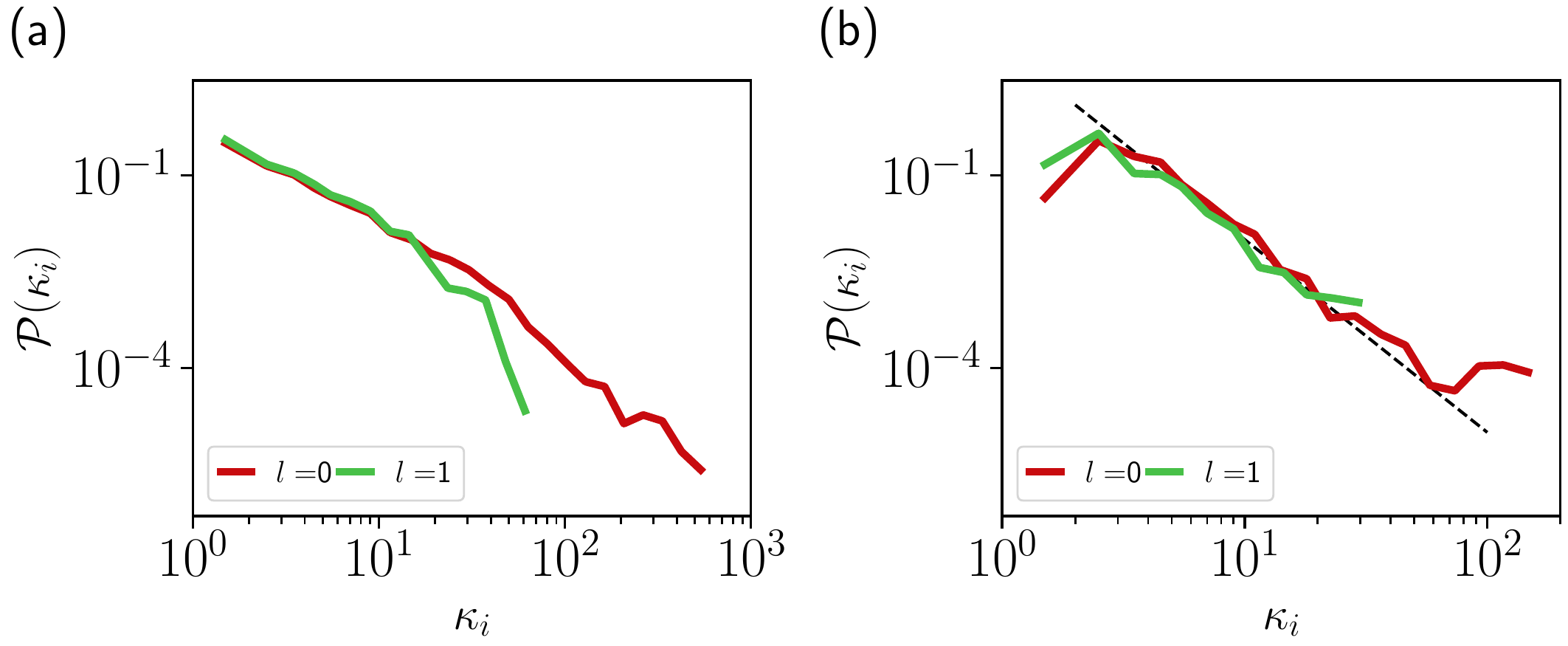}
 \caption{RG transformation of real networks. Degree distribution after a RG step for: (a) $A.~Thaliana$ using $\tau=1.0$ and (b) $D.~Melanogaster$ metabolic network using $\tau=0.1$.}\label{RWorld}
 \end{center}
\end{figure}

Finally, we apply our RG approach to different scale-free real networks, which have been demonstrated to follow a finite-size scaling hypothesis without any self-tuning \cite{Serafino2021}, producing different downscaled network replicas. Figure \ref{RWorld} shows the particular case of the structure of metabolic networks: $A.~Thaliana$ \cite{das2012} and $D.~Melanogaster$ \cite{Dros} (see also SI5 \cite{SI}), thus confirming that these networks are inherently scale-free.
%CONCLUSIONS

%Figure \ref{RWorld} (b) shows ....
\section*{Laplacian Renormalization Group (LRG)}
Thus far, as in the Kadanoff hypothesis, we do not have a shred of justification for our assumptions. In this section, we introduce a rigorous formulation of the Laplacian RG for undirected graph with $N$ nodes, which can be appropriately seen as the analogous of the field theory $k-$space RG {\em \`a la Wilson} in statistical physics. From this formulation, we get a natural definition of the Kadanoff blocks of nodes at each step of renormalization, of which the definition given in the previous section can be seen as a practical and useful approximation.

Let us consider the case in which we want to renormalize the information diffusion on the graph up to a time $\tau^*$ so that to keep only diffusion modes on scale larger than $\tau^*$ (e.g. where $C$ shows a maximum). In order to due this it is convenient to adopt the bra-ket formalism in which $\bra{i}\ket{\lambda}$ indicate the component of the Laplacian eigenvector $\ket{\lambda}$ due to the $i^{\text{th}}$ node of the graph (in this sense we can identify $\ket{i}$ with the normalized $N-$dimensional column-vector of all components $0$ with the exception of the $i^{\text{th}}$ which is $1$). In this notation the Laplacian operator is $\sum_{\lambda}\lambda\ket{\lambda}\!\bra{\lambda}$ and we identify the $n<N$ eigenvalues $\lambda\ge =\lambda^*=1/\tau^*$ and the related eigenvectors $\ket{\lambda}$. An RG step consists in integrating out these diffusion eigenmodes from the Laplacian and appropriately rescaling the graph, namely:\\
(i) Reduce the Laplacian operator to the contribution of the $N-n$ slow eigenvectors with $\lambda<\lambda^*$, $\hat L'=\sum_{\lambda<\lambda^*}\lambda \ket{\lambda}\!\bra{\lambda}$;\\
(ii) Subsequently, form $N-n$ macronodes from the $N$ original graph nodes by progressively aggregating them in $N-n$ clusters and identifying each cluster with a macronode. This is obtained by connecting the nodes in clusters following the values of $|\rho_{ij}(\tau^*)|=|\bra{i}\hat \rho(\tau^*)\ket{j}|$ in descending order and stopping when $N-n$ clusters are finally obtained;\\
(iii) The $(N-n)\times(N-n)$ weighted adjacency matrix $\hat A'$ giving the connections between macronodes is then simply obtained by putting $A'_{\alpha\beta}=-L'_{\alpha\beta}=-\bra{\alpha}\hat L'\ket{\beta}$, for $\alpha\ne \beta$, where $\ket{\alpha}$ (and $\ket{\beta}$) is the $N-$dimensional normalized column ket given by the superposition of the $\ket{i}$ representing the nodes of the original graph aggregated in the macronode $\alpha$, i.e. the $N-$dimensional column vector with unitary components corresponding to all the nodes aggregated in the macronode and zero otherwise. Moreover, we set $A'_{\alpha\alpha}=0$ and $L'_{\alpha\alpha}=\sum_{\beta}A'_{\alpha\beta}$. \\
(iv) Finally, as in usual RG procedures, we have to rescale time $t\to t'$, so that $\tau^*$ in $t$ becomes the unitary interval in the rescaled time variable $t'$. This is done by defining $t'=t/\tau^*$ and, consequently, redefining the coarse-grained Laplacian operator as $\hat L''=\tau^*\hat L'$.

In this way, we have defined a consistent Laplacian-driven renormalization step of the graph, reducing the dimension from $N$ to $N-n$. 
It is crucial to notice that, as defined above, this formulation of the  RG is exactly the extension to graphs of the $k-$space RG defined in statistical mechanics. Indeed, in metric spaces, the Laplacian operator has eigenvalues proportional to $k^2$, where $k$ are the wave vectors of the modes, and the corresponding eigenvector is the plane waves with wave vector $k$. In addition, the correspondence of the operator $\hat \rho(\tau)$ with the Boltzmann factor $e^{-\beta \hat H}$ in statistical field theory makes our method strictly corresponding to the renormalization of the free field theory. 
Finally, note that even though we start with a binary graph, we end up with a weighted full one. In order to visualize better the resulting graph of macronodes, a reasonable decimation recipe --tantamount to the majority rule-- can then be adopted to get a binary graph again. We pinpoint that the particular election of $\tau^*$ ensures the maximal iterability of the LRG scheme, even if any other election can generate alternative network reductions for a few RG steps. We must emphasize that as $\tau^*$ identifies points of fast information diffusion across the network, identifying large $\tau$ values to perform Kadanoff supernodes (or to integrate many network modes) will produce a drastic network reduction.

\section*{Conclusions}
RG represents a significant development in contemporary statistical mechanics \cite{Binney, Amit, Kardar}. Their application to diverse dynamical processes operating on top of regular spatial structures (i.e., lattices) allows the introduction of the idea of \emph{universality} and the classification of models (otherwise presumed faraway)  within a small number of universality classes. Examples run, from ferromagnetic systems \cite{Kadanoff1966} to percolation \cite{Harris1975}, polymers, and particle physics \cite{DeGennes1979}. Recently, groundbreaking applications have been addressed the problem in complex biological systems \cite{PVilla} illuminating collective behavior of neurons in mouse hippocampus \cite{meshulam2019}, or dynamical couplings in natural swarms \cite{Cavagna2019}.

There is no apparent equivalence to analyzing RG processes in complex spatial structures, even if some pioneering approaches have recently proposed sound procedures to state equivalent general RG schemes to those of statistical physics. The most promising approaches draw on \emph{hidden metric} assumptions, spatially mapping nodes in some abstract topological space, which must be considered as an 'a priori' hypothesis \cite{Garcia2018, Zheng2020}. Despite that, they show fundamental problems to maintain intrinsic network properties \cite{Garcia2018, Gfeller} --e.g., connectivity-- of reduced replicas when performing decimation. Moreover, setting the equivalent to conventional RG flow without any spatial projection of the nodes or grouping premises \cite{Garlaschelli}, but induced by diffusion distances \cite{Boguna2021} remains an unsolved fundamental problem.

We here develop an RG scheme based on information diffusion distances which generate an effective defocusing of the network, allowing us to select, through the analysis of the points with maximal entropy' change, the critical resolution scales of the network. By linking them with new mesoscopic scales, we can first perform the corresponding Kadanoff real-space renormalization in complex networks, solving particular problems such as decimation. To perform the real-space dynamical RG scheme, two fundamental scales must be considered: the lattice space ($a$) and the correlation length of the system ($\xi$). In concomitance with the original formulation, the peaks in the specific heat of the information diffusion flow allow us to identify characteristic scales of the system. They, therefore, represent the counterpart to the correlation length or the lattice spacing when the process is carried out over spins' blocks or active sites in percolation phenomena \cite{Moloney}. We point out that complex networks exhibit degree heterogeneity, thus lacking indistinguishable groups of nodes under the Kadanoff blocking scheme, i.e., the blocks need to reflect the intrinsic heterogeneous architecture of the network. This conundrum automatically leads to many possibilities in grouping nodes, making the problem looks unaffordable.

The study of the Laplacian spectrum allows us to compute the network modes, playing the exact role of $a$ and $\Lambda$ in RG schemes (i.e., the analogy with the UV cutoff \cite{ZinnJustin, Binney}). We point out that for selecting critical scales of the network, all eigenvalues (fluctuations) can be of utmost relevance to give us information about the intertwined network scales, relating the short-distance cutoff and the macroscopic scale. In particular, our framework conserves the main network properties \cite{Gfeller} --e.g., the average degree \cite{Garcia2018}-- of the downscaled networks (only if they are truly scale-invariant \cite{Serafino2021}), thus confirming the existence of repulsive, non-trivial, fixed points in the RG flow. It also allows us to extract mesoscopic information of networks communities even if they are blatantly scale-dependent. Our RG scheme solves a crucial open problem \cite{Garlaschelli}: the limited iterability in small-world networks due to short path lengths, limited only by the network size.

Altogether, we propose here a new RG approach \emph{\`a la Wilson} --therefore working in the momentum space-- based on the Laplacian properties of the network: the Laplacian Renormalization Group (LRG), which strictly corresponds to the renormalization of the free field theory in network structures. We also demonstrate that only true scale-invariant networks will exhibit constant specific-heat values at all resolution scales, even if it is possible to define weak scale-invariant structures that can be renormalized anyway. Barabasi-Albert networks which present ultra-small world properties \cite{Cohen2003} are a prime example of this (see SI4 \cite{SI}), being, therefore, the equivalent to the Watts-Strogatz process to lattice systems \cite{WS}.

Our RG scheme opens a route to extend RG flow study further, developing a common mathematical framework classifying complex networks in universality classes. Finally, subsequent perturbative analyses can also help in shedding light on co-existing temporal scales due to the interplay between structure and dynamics.

\section*{Acknowledgments}

 P.V. acknowledge financial support from the Spanish "Ministerio de Ciencia e Innovación” and the "Agencia Estatal de Investigación (AEI)” under Project Ref. PID2020-113681GB-I00. We also thank G. Cimini and D. Garlaschelli for extremely valuable suggestions on earlier versions of the manuscript.

\section*{Methods}

\subsection{Statistical physics of information network diffusion}

Let us consider the adjacency matrix of a simple binary graph, $\hat A$, and define $\hat L= \hat D - \hat A$ as the 'fluid Laplacian matrix' \cite{Masuda2017}, where $D_{ij}=k_i\delta_{ij}$ and $k_i$ is the mean connectivity of each node. In terms of the network propagator, $\hat K=e^{-\tau \hat L}$, it is possible to define the ensemble of accesible information diffusion states \cite{Domenico2016,ghavasieh2020,Villegas2022}, namely,
\begin{equation}
 \mathbf{\rho(\tau)}= \frac{e^{-\tau \hat L}}{Z}\,,
 \label{EvolMat}
\end{equation}
where $\rho(\tau)$ is tantamount to the canonical density operator in statistical physics (or to the functional over fields configurations) \cite{Binney,Pathria,Greiner}, and $Z=\stackrel[i=1]{N}{\sum}e^{-\lambda_i}$, being $\lambda_i$ the set of system' eigenvalues. Since for a simple graph $\hat L$ is a Hermitian matrix, $\hat L$ plays the role of the Hamiltonian operator and $\tau$ the role of the inverse temperature. It is possible to therefore define the network entropy \cite{Domenico2016} through the relation
%\begin{multline}
% S[\rho(\tau)]=-tr[\hat \rho(\tau) \log \hat \rho(\tau)]=\frac{\tau}{Z}tr[\hat L e^{-\tau \hat L}+ e^{-\tau \hat L}\log Z]\\=\tau[\langle \hat L \rangle_t + \log Z]
%\end{multline}
\begin{multline}
 S[\rho(\tau)]=-Tr\left[\hat \rho(\tau) \log \hat \rho(\tau)\right]=Tr\left[\frac{e^{-\tau \hat L}}{Z}(\tau \hat L + \log Z)\right]\\ = \tau \langle \lambda \rangle_t + \log Z
 \end{multline}

being $\langle \hat O \rangle_t=Tr[ \hat \rho \hat O ]$. Immediately, it is possible to define the specific heat of the network as,

%\begin{equation}
% C(\tau)=\tau \frac{dS}{d\tau}=\tau[\langle \hat L \rangle_t + \log Z]+\tau^2 [\frac{d\langle \hat L \rangle_t}{dt}+\frac{dZ/dt}{Z(\tau)}]
%end{equation}

\begin{equation}
 C=-\frac{dS}{d \log \tau}=-\tau^2\frac{d\langle \lambda \rangle_t}{d\tau}
 \label{SHeat}
\end{equation}

\subsection{Informational phase transitions}

The specific heat of the network of Eq.\eqref{SHeat} is a detector of transition points corresponding to the intrinsic characteristic diffusion scales of the network. In particular, the condition $\ensuremath{\left.\frac{dC}{d\tau}\right|_{\tau^{*}}=0}$ defines $\tau^*$, and reveals the existence of pronounced peaks revealing a strong deceleration of the information diffusion. Moreover, employing the thermal fluctuation-dissipation theorem \cite{Marro, Moloney} the specific heat links to entropy fluctuations making $C$ proportional to $\sigma^2_S=\langle S^2 \rangle - \langle S \rangle^2$ which, over many independent realizations, we expect scales as $1/N$ where $N$ is the number of nodes of the network (as a direct application of the central limit theorem \cite{Gardiner}).

\subsection{Scale-invariant networks}
Let us now define {\em informationally} scale-invariant networks in agreement with our definition of the Laplacian RG. A network has scale-invariant properties in a resolution region, if the entropic susceptibility/specific heat $C$ takes a constant value $C_1>0$ in the corresponding diffusion time interval. This property indeed describes a situation in which the informational entropy increases of the same amount in two equal logarithmic time scales, which means a scale-invariant transmission of information at every network resolution scale. 
It is matter of simple algebra to see, from Eq~\eqref{SHeat}, to see that this means
\[\frac{d\langle \lambda \rangle_\tau}{d\tau}=-\frac{C_1}{\tau^2}\,.\]
Knowing that for $\tau\rightarrow\infty$, $\langle \lambda \rangle \rightarrow 0$, by integration we find 
\[\langle \lambda \rangle_\tau=\frac{C_1}{\tau}=\nicefrac{\stackrel[i=1]{N}{\sum}\lambda_i e^{-\tau\lambda_i}}{\stackrel[i=1]{N}{\sum}e^{-\tau\lambda_i}}\,.\] 
If we now assume the continuum approximation of the Laplacian spectrum and the general condition $P(\lambda) \sim \lambda^{\gamma}$ for the informational scale-invariance, we finally found the following relation between the constant values of the specific heat $C(\tau)$ and the scale invariance exponent of $P(\lambda)$ (for large diffusion times, i.e. small $\lambda$):
\begin{equation}
    C_1=\frac{\Gamma(\gamma+2)}{\Gamma(\gamma+1)}=\gamma+1,
\end{equation}
where $\Gamma(z)$ is the Euler's gamma function.

%\def\url#1{}
%\bibliography{Entropy} 

\begin{thebibliography}{48}%
\makeatletter
\providecommand \@ifxundefined [1]{%
 \@ifx{#1\undefined}
}%
\providecommand \@ifnum [1]{%
 \ifnum #1\expandafter \@firstoftwo
 \else \expandafter \@secondoftwo
 \fi
}%
\providecommand \@ifx [1]{%
 \ifx #1\expandafter \@firstoftwo
 \else \expandafter \@secondoftwo
 \fi
}%
\providecommand \natexlab [1]{#1}%
\providecommand \enquote  [1]{``#1''}%
\providecommand \bibnamefont  [1]{#1}%
\providecommand \bibfnamefont [1]{#1}%
\providecommand \citenamefont [1]{#1}%
\providecommand \href@noop [0]{\@secondoftwo}%
\providecommand \href [0]{\begingroup \@sanitize@url \@href}%
\providecommand \@href[1]{\@@startlink{#1}\@@href}%
\providecommand \@@href[1]{\endgroup#1\@@endlink}%
\providecommand \@sanitize@url [0]{\catcode `\\12\catcode `\$12\catcode
  `\&12\catcode `\#12\catcode `\^12\catcode `\_12\catcode `\%12\relax}%
\providecommand \@@startlink[1]{}%
\providecommand \@@endlink[0]{}%
\providecommand \url  [0]{\begingroup\@sanitize@url \@url }%
\providecommand \@url [1]{\endgroup\@href {#1}{\urlprefix }}%
\providecommand \urlprefix  [0]{URL }%
\providecommand \Eprint [0]{\href }%
\providecommand \doibase [0]{https://doi.org/}%
\providecommand \selectlanguage [0]{\@gobble}%
\providecommand \bibinfo  [0]{\@secondoftwo}%
\providecommand \bibfield  [0]{\@secondoftwo}%
\providecommand \translation [1]{[#1]}%
\providecommand \BibitemOpen [0]{}%
\providecommand \bibitemStop [0]{}%
\providecommand \bibitemNoStop [0]{.\EOS\space}%
\providecommand \EOS [0]{\spacefactor3000\relax}%
\providecommand \BibitemShut  [1]{\csname bibitem#1\endcsname}%
\let\auto@bib@innerbib\@empty
%</preamble>
\bibitem [{\citenamefont {Fisher}(1974)}]{Fisher1974}%
  \BibitemOpen
  \bibfield  {author} {\bibinfo {author} {\bibfnamefont {M.~E.}\ \bibnamefont
  {Fisher}},\ }\bibfield  {title} {\bibinfo {title} {The renormalization group
  in the theory of critical behavior},\ }\href
  {https://doi.org/10.1103/RevModPhys.46.597} {\bibfield  {journal} {\bibinfo
  {journal} {Rev. Mod. Phys.}\ }\textbf {\bibinfo {volume} {46}},\ \bibinfo
  {pages} {597} (\bibinfo {year} {1974})}\BibitemShut {NoStop}%
\bibitem [{\citenamefont {Wilson}\ and\ \citenamefont
  {Kogut}(1974)}]{Wilson1974}%
  \BibitemOpen
  \bibfield  {author} {\bibinfo {author} {\bibfnamefont {K.~G.}\ \bibnamefont
  {Wilson}}\ and\ \bibinfo {author} {\bibfnamefont {J.}~\bibnamefont {Kogut}},\
  }\bibfield  {title} {\bibinfo {title} {The renormalization group and the
  $\epsilon$ expansion},\ }\href {https://doi.org/10.1016/0370-1573(74)90023-4}
  {\bibfield  {journal} {\bibinfo  {journal} {Phys. Rep.}\ }\textbf {\bibinfo
  {volume} {12}},\ \bibinfo {pages} {75} (\bibinfo {year} {1974})}\BibitemShut
  {NoStop}%
\bibitem [{\citenamefont {Binney}\ \emph {et~al.}(1992)\citenamefont {Binney},
  \citenamefont {Dowrick}, \citenamefont {Fisher},\ and\ \citenamefont
  {Newman}}]{Binney}%
  \BibitemOpen
  \bibfield  {author} {\bibinfo {author} {\bibfnamefont {J.~J.}\ \bibnamefont
  {Binney}}, \bibinfo {author} {\bibfnamefont {N.~J.}\ \bibnamefont {Dowrick}},
  \bibinfo {author} {\bibfnamefont {A.~J.}\ \bibnamefont {Fisher}},\ and\
  \bibinfo {author} {\bibfnamefont {M.~E.}\ \bibnamefont {Newman}},\
  }\href@noop {} {\emph {\bibinfo {title} {The theory of critical phenomena: an
  introduction to the renormalization group}}}\ (\bibinfo  {publisher} {Oxford
  University Press},\ \bibinfo {address} {Oxford},\ \bibinfo {year}
  {1992})\BibitemShut {NoStop}%
\bibitem [{\citenamefont {Amit}\ and\ \citenamefont
  {Martin-Mayor}(2005)}]{Amit}%
  \BibitemOpen
  \bibfield  {author} {\bibinfo {author} {\bibfnamefont {D.~J.}\ \bibnamefont
  {Amit}}\ and\ \bibinfo {author} {\bibfnamefont {V.}~\bibnamefont
  {Martin-Mayor}},\ }\href {https://doi.org/10.1142/5715} {\emph {\bibinfo
  {title} {Field Theory, the Renormalization Group, and Critical Phenomena}}},\
  \bibinfo {edition} {3rd}\ ed.\ (\bibinfo  {publisher} {World Scientific},\
  \bibinfo {address} {Singapore},\ \bibinfo {year} {2005})\BibitemShut
  {NoStop}%
\bibitem [{\citenamefont {Kardar}(2007)}]{Kardar}%
  \BibitemOpen
  \bibfield  {author} {\bibinfo {author} {\bibfnamefont {M.}~\bibnamefont
  {Kardar}},\ }\href {https://doi.org/10.1017/CBO9780511815881} {\emph
  {\bibinfo {title} {Statistical physics of fields}}}\ (\bibinfo  {publisher}
  {Cambridge University Press},\ \bibinfo {address} {Cambridge},\ \bibinfo
  {year} {2007})\BibitemShut {NoStop}%
\bibitem [{\citenamefont {Gfeller}\ and\ \citenamefont
  {De~Los~Rios}(2007)}]{Gfeller}%
  \BibitemOpen
  \bibfield  {author} {\bibinfo {author} {\bibfnamefont {D.}~\bibnamefont
  {Gfeller}}\ and\ \bibinfo {author} {\bibfnamefont {P.}~\bibnamefont
  {De~Los~Rios}},\ }\bibfield  {title} {\bibinfo {title} {Spectral coarse
  graining of complex networks},\ }\href
  {https://doi.org/10.1103/PhysRevLett.99.038701} {\bibfield  {journal}
  {\bibinfo  {journal} {Phys. Rev. Lett.}\ }\textbf {\bibinfo {volume} {99}},\
  \bibinfo {pages} {038701} (\bibinfo {year} {2007})}\BibitemShut {NoStop}%
\bibitem [{\citenamefont {Song}\ \emph {et~al.}(2005)\citenamefont {Song},
  \citenamefont {Havlin},\ and\ \citenamefont {Makse}}]{Song2005}%
  \BibitemOpen
  \bibfield  {author} {\bibinfo {author} {\bibfnamefont {C.}~\bibnamefont
  {Song}}, \bibinfo {author} {\bibfnamefont {S.}~\bibnamefont {Havlin}},\ and\
  \bibinfo {author} {\bibfnamefont {H.~A.}\ \bibnamefont {Makse}},\ }\bibfield
  {title} {\bibinfo {title} {Self-similarity of complex networks},\ }\href
  {https://doi.org/10.1038/nature03248} {\bibfield  {journal} {\bibinfo
  {journal} {Nature}\ }\textbf {\bibinfo {volume} {433}},\ \bibinfo {pages}
  {392} (\bibinfo {year} {2005})}\BibitemShut {NoStop}%
\bibitem [{\citenamefont {Song}\ \emph {et~al.}(2006)\citenamefont {Song},
  \citenamefont {Havlin},\ and\ \citenamefont {Makse}}]{Song2006}%
  \BibitemOpen
  \bibfield  {author} {\bibinfo {author} {\bibfnamefont {C.}~\bibnamefont
  {Song}}, \bibinfo {author} {\bibfnamefont {S.}~\bibnamefont {Havlin}},\ and\
  \bibinfo {author} {\bibfnamefont {H.~A.}\ \bibnamefont {Makse}},\ }\bibfield
  {title} {\bibinfo {title} {Origins of fractality in the growth of complex
  networks},\ }\href {https://doi.org/10.1038/nphys266} {\bibfield  {journal}
  {\bibinfo  {journal} {Nat. Phys.}\ }\textbf {\bibinfo {volume} {2}},\
  \bibinfo {pages} {275} (\bibinfo {year} {2006})}\BibitemShut {NoStop}%
\bibitem [{\citenamefont {Goh}\ \emph {et~al.}(2006)\citenamefont {Goh},
  \citenamefont {Salvi}, \citenamefont {Kahng},\ and\ \citenamefont
  {Kim}}]{Goh2006}%
  \BibitemOpen
  \bibfield  {author} {\bibinfo {author} {\bibfnamefont {K.-I.}\ \bibnamefont
  {Goh}}, \bibinfo {author} {\bibfnamefont {G.}~\bibnamefont {Salvi}}, \bibinfo
  {author} {\bibfnamefont {B.}~\bibnamefont {Kahng}},\ and\ \bibinfo {author}
  {\bibfnamefont {D.}~\bibnamefont {Kim}},\ }\bibfield  {title} {\bibinfo
  {title} {Skeleton and fractal scaling in complex networks},\ }\href
  {https://doi.org/10.1103/PhysRevLett.96.018701} {\bibfield  {journal}
  {\bibinfo  {journal} {Phys. Rev. Lett.}\ }\textbf {\bibinfo {volume} {96}},\
  \bibinfo {pages} {018701} (\bibinfo {year} {2006})}\BibitemShut {NoStop}%
\bibitem [{\citenamefont {Kim}\ \emph {et~al.}(2007)\citenamefont {Kim},
  \citenamefont {Goh}, \citenamefont {Kahng},\ and\ \citenamefont
  {Kim}}]{Kim2007}%
  \BibitemOpen
  \bibfield  {author} {\bibinfo {author} {\bibfnamefont {J.~S.}\ \bibnamefont
  {Kim}}, \bibinfo {author} {\bibfnamefont {K.-I.}\ \bibnamefont {Goh}},
  \bibinfo {author} {\bibfnamefont {B.}~\bibnamefont {Kahng}},\ and\ \bibinfo
  {author} {\bibfnamefont {D.}~\bibnamefont {Kim}},\ }\bibfield  {title}
  {\bibinfo {title} {Fractality and self-similarity in scale-free networks},\
  }\href {https://doi.org/10.1088/1367-2630/9/6/177} {\bibfield  {journal}
  {\bibinfo  {journal} {New J. Phys.}\ }\textbf {\bibinfo {volume} {9}},\
  \bibinfo {pages} {177} (\bibinfo {year} {2007})}\BibitemShut {NoStop}%
\bibitem [{\citenamefont {Radicchi}\ \emph {et~al.}(2008)\citenamefont
  {Radicchi}, \citenamefont {Ramasco}, \citenamefont {Barrat},\ and\
  \citenamefont {Fortunato}}]{Radicchi2008}%
  \BibitemOpen
  \bibfield  {author} {\bibinfo {author} {\bibfnamefont {F.}~\bibnamefont
  {Radicchi}}, \bibinfo {author} {\bibfnamefont {J.~J.}\ \bibnamefont
  {Ramasco}}, \bibinfo {author} {\bibfnamefont {A.}~\bibnamefont {Barrat}},\
  and\ \bibinfo {author} {\bibfnamefont {S.}~\bibnamefont {Fortunato}},\
  }\bibfield  {title} {\bibinfo {title} {Complex networks renormalization:
  Flows and fixed points},\ }\href
  {https://doi.org/10.1103/PhysRevLett.101.148701} {\bibfield  {journal}
  {\bibinfo  {journal} {Phys. Rev. Lett.}\ }\textbf {\bibinfo {volume} {101}},\
  \bibinfo {pages} {148701} (\bibinfo {year} {2008})}\BibitemShut {NoStop}%
\bibitem [{\citenamefont {Rozenfeld}\ \emph {et~al.}(2010)\citenamefont
  {Rozenfeld}, \citenamefont {Song},\ and\ \citenamefont
  {Makse}}]{Rozenfeld2010}%
  \BibitemOpen
  \bibfield  {author} {\bibinfo {author} {\bibfnamefont {H.~D.}\ \bibnamefont
  {Rozenfeld}}, \bibinfo {author} {\bibfnamefont {C.}~\bibnamefont {Song}},\
  and\ \bibinfo {author} {\bibfnamefont {H.~A.}\ \bibnamefont {Makse}},\
  }\bibfield  {title} {\bibinfo {title} {Small-world to fractal transition in
  complex networks: a renormalization group approach},\ }\href
  {https://doi.org/10.1103/PhysRevLett.104.025701} {\bibfield  {journal}
  {\bibinfo  {journal} {Phys. Rev. Lett.}\ }\textbf {\bibinfo {volume} {104}},\
  \bibinfo {pages} {025701} (\bibinfo {year} {2010})}\BibitemShut {NoStop}%
\bibitem [{\citenamefont {Watts}\ and\ \citenamefont {Strogatz}(1998)}]{WS}%
  \BibitemOpen
  \bibfield  {author} {\bibinfo {author} {\bibfnamefont {D.~J.}\ \bibnamefont
  {Watts}}\ and\ \bibinfo {author} {\bibfnamefont {S.~H.}\ \bibnamefont
  {Strogatz}},\ }\bibfield  {title} {\bibinfo {title} {Collective dynamics of
  ‘small-world’networks},\ }\href {https://doi.org/10.1038/30918}
  {\bibfield  {journal} {\bibinfo  {journal} {Nature}\ }\textbf {\bibinfo
  {volume} {393}},\ \bibinfo {pages} {440} (\bibinfo {year}
  {1998})}\BibitemShut {NoStop}%
\bibitem [{\citenamefont {Klemm}\ and\ \citenamefont
  {Eguiluz}(2002)}]{Klemm2002}%
  \BibitemOpen
  \bibfield  {author} {\bibinfo {author} {\bibfnamefont {K.}~\bibnamefont
  {Klemm}}\ and\ \bibinfo {author} {\bibfnamefont {V.~M.}\ \bibnamefont
  {Eguiluz}},\ }\bibfield  {title} {\bibinfo {title} {Growing scale-free
  networks with small-world behavior},\ }\href
  {https://doi.org/10.1103/PhysRevE.65.057102} {\bibfield  {journal} {\bibinfo
  {journal} {Phys. Rev. E}\ }\textbf {\bibinfo {volume} {65}},\ \bibinfo
  {pages} {057102} (\bibinfo {year} {2002})}\BibitemShut {NoStop}%
\bibitem [{\citenamefont {Garuccio}\ \emph {et~al.}()\citenamefont {Garuccio},
  \citenamefont {Lalli},\ and\ \citenamefont {Garlaschelli}}]{Garlaschelli}%
  \BibitemOpen
  \bibfield  {author} {\bibinfo {author} {\bibfnamefont {E.}~\bibnamefont
  {Garuccio}}, \bibinfo {author} {\bibfnamefont {M.}~\bibnamefont {Lalli}},\
  and\ \bibinfo {author} {\bibfnamefont {D.}~\bibnamefont {Garlaschelli}},\
  }\bibfield  {title} {\bibinfo {title} {Multiscale network renormalization:
  scale-invariance without geometry},\ }\href
  {https://arxiv.org/abs/2009.11024} {\bibinfo  {journal} {arXiv preprint
  arXiv:2009.11024}\ }\BibitemShut {NoStop}%
\bibitem [{\citenamefont {Garc{\'\i}a-P{\'e}rez}\ \emph
  {et~al.}(2018)\citenamefont {Garc{\'\i}a-P{\'e}rez}, \citenamefont
  {Bogu{\~n}{\'a}},\ and\ \citenamefont {Serrano}}]{Garcia2018}%
  \BibitemOpen
\bibfield  {journal} {  }\bibfield  {author} {\bibinfo {author} {\bibfnamefont
  {G.}~\bibnamefont {Garc{\'\i}a-P{\'e}rez}}, \bibinfo {author} {\bibfnamefont
  {M.}~\bibnamefont {Bogu{\~n}{\'a}}},\ and\ \bibinfo {author} {\bibfnamefont
  {M.~{\'A}.}\ \bibnamefont {Serrano}},\ }\bibfield  {title} {\bibinfo {title}
  {Multiscale unfolding of real networks by geometric renormalization},\ }\href
  {https://doi.org/10.1038/s41567-018-0072-5} {\bibfield  {journal} {\bibinfo
  {journal} {Nat. Phys.}\ }\textbf {\bibinfo {volume} {14}},\ \bibinfo {pages}
  {583} (\bibinfo {year} {2018})}\BibitemShut {NoStop}%
\bibitem [{\citenamefont {Zheng}\ \emph {et~al.}(2020)\citenamefont {Zheng},
  \citenamefont {Allard}, \citenamefont {Hagmann}, \citenamefont
  {Alem{\'a}n-G{\'o}mez},\ and\ \citenamefont {Serrano}}]{Zheng2020}%
  \BibitemOpen
  \bibfield  {author} {\bibinfo {author} {\bibfnamefont {M.}~\bibnamefont
  {Zheng}}, \bibinfo {author} {\bibfnamefont {A.}~\bibnamefont {Allard}},
  \bibinfo {author} {\bibfnamefont {P.}~\bibnamefont {Hagmann}}, \bibinfo
  {author} {\bibfnamefont {Y.}~\bibnamefont {Alem{\'a}n-G{\'o}mez}},\ and\
  \bibinfo {author} {\bibfnamefont {M.~{\'A}.}\ \bibnamefont {Serrano}},\
  }\bibfield  {title} {\bibinfo {title} {Geometric renormalization unravels
  self-similarity of the multiscale human connectome},\ }\href
  {https://doi.org/10.1073/pnas.1922248117} {\bibfield  {journal} {\bibinfo
  {journal} {Proc. Natl. Acad. Sci. U.S.A}\ }\textbf {\bibinfo {volume}
  {117}},\ \bibinfo {pages} {20244} (\bibinfo {year} {2020})}\BibitemShut
  {NoStop}%
\bibitem [{\citenamefont {Bogu{\~n}{\'a}}\ \emph {et~al.}(2021)\citenamefont
  {Bogu{\~n}{\'a}}, \citenamefont {Bonamassa}, \citenamefont {De~Domenico},
  \citenamefont {Havlin}, \citenamefont {Krioukov},\ and\ \citenamefont
  {Serrano}}]{Boguna2021}%
  \BibitemOpen
  \bibfield  {author} {\bibinfo {author} {\bibfnamefont {M.}~\bibnamefont
  {Bogu{\~n}{\'a}}}, \bibinfo {author} {\bibfnamefont {I.}~\bibnamefont
  {Bonamassa}}, \bibinfo {author} {\bibfnamefont {M.}~\bibnamefont
  {De~Domenico}}, \bibinfo {author} {\bibfnamefont {S.}~\bibnamefont {Havlin}},
  \bibinfo {author} {\bibfnamefont {D.}~\bibnamefont {Krioukov}},\ and\
  \bibinfo {author} {\bibfnamefont {M.}~\bibnamefont {Serrano}},\ }\bibfield
  {title} {\bibinfo {title} {Network geometry},\ }\href
  {https://doi.org/10.1038/s42254-020-00264-4} {\bibfield  {journal} {\bibinfo
  {journal} {Nat. Rev. Phys.}\ }\textbf {\bibinfo {volume} {3}},\ \bibinfo
  {pages} {114} (\bibinfo {year} {2021})}\BibitemShut {NoStop}%
\bibitem [{\citenamefont {Kosterlitz}\ and\ \citenamefont
  {Thouless}()}]{Kosterlitz}%
  \BibitemOpen
  \bibfield  {author} {\bibinfo {author} {\bibfnamefont {J.~M.}\ \bibnamefont
  {Kosterlitz}}\ and\ \bibinfo {author} {\bibfnamefont {D.~J.}\ \bibnamefont
  {Thouless}},\ }\bibinfo {title} {Early work on defect driven phase
  transitions},\ in\ \href {https://doi.org/10.1142/9789814417648_0001} {\emph
  {\bibinfo {booktitle} {40 Years of Berezinskii–Kosterlitz–Thouless
  Theory}}},\ pp.\ \bibinfo {pages} {1--67}\BibitemShut {NoStop}%
\bibitem [{\citenamefont {Zinn-Justin}(2007)}]{ZinnJustin}%
  \BibitemOpen
  \bibfield  {author} {\bibinfo {author} {\bibfnamefont {J.}~\bibnamefont
  {Zinn-Justin}},\ }\href@noop {} {\emph {\bibinfo {title} {Phase transitions
  and renormalization group}}}\ (\bibinfo  {publisher} {Oxford University Press
  on Demand},\ \bibinfo {year} {2007})\BibitemShut {NoStop}%
\bibitem [{\citenamefont {Gardner}\ \emph {et~al.}(1984)\citenamefont
  {Gardner}, \citenamefont {Itzykson},\ and\ \citenamefont
  {Derrida}}]{Gardner}%
  \BibitemOpen
  \bibfield  {author} {\bibinfo {author} {\bibfnamefont {E.}~\bibnamefont
  {Gardner}}, \bibinfo {author} {\bibfnamefont {C.}~\bibnamefont {Itzykson}},\
  and\ \bibinfo {author} {\bibfnamefont {B.}~\bibnamefont {Derrida}},\
  }\bibfield  {title} {\bibinfo {title} {The laplacian on a random
  one-dimensional lattice},\ }\href
  {https://doi.org/10.1088/0305-4470/17/5/030} {\bibfield  {journal} {\bibinfo
  {journal} {J. Phys. A Math. Gen.}\ }\textbf {\bibinfo {volume} {17}},\
  \bibinfo {pages} {1093} (\bibinfo {year} {1984})}\BibitemShut {NoStop}%
\bibitem [{\citenamefont {Kadanoff}(1966)}]{Kadanoff1966}%
  \BibitemOpen
  \bibfield  {author} {\bibinfo {author} {\bibfnamefont {L.~P.}\ \bibnamefont
  {Kadanoff}},\ }\bibfield  {title} {\bibinfo {title} {Scaling laws for ising
  models near $t_c$},\ }\href
  {https://doi.org/10.1103/PhysicsPhysiqueFizika.2.263} {\bibfield  {journal}
  {\bibinfo  {journal} {Phys. Phys. Fiz.}\ }\textbf {\bibinfo {volume} {2}},\
  \bibinfo {pages} {263} (\bibinfo {year} {1966})}\BibitemShut {NoStop}%
\bibitem [{\citenamefont {Matsumoto}\ \emph {et~al.}(2020)\citenamefont
  {Matsumoto}, \citenamefont {Tanaka},\ and\ \citenamefont
  {Tsuchiya}}]{Matsumoto2020}%
  \BibitemOpen
  \bibfield  {author} {\bibinfo {author} {\bibfnamefont {M.}~\bibnamefont
  {Matsumoto}}, \bibinfo {author} {\bibfnamefont {G.}~\bibnamefont {Tanaka}},\
  and\ \bibinfo {author} {\bibfnamefont {A.}~\bibnamefont {Tsuchiya}},\
  }\bibfield  {title} {\bibinfo {title} {The renormalization group and the
  diffusion equation},\ }\bibfield  {journal} {\bibinfo  {journal} {Prog.
  Theor. Exp. Phys.}\ }\textbf {\bibinfo {volume} {2021}},\ \href
  {https://doi.org/10.1093/ptep/ptaa175} {10.1093/ptep/ptaa175} (\bibinfo
  {year} {2020})\BibitemShut {NoStop}%
\bibitem [{\citenamefont {Masuda}\ \emph {et~al.}(2017)\citenamefont {Masuda},
  \citenamefont {Porter},\ and\ \citenamefont {Lambiotte}}]{Masuda2017}%
  \BibitemOpen
  \bibfield  {author} {\bibinfo {author} {\bibfnamefont {N.}~\bibnamefont
  {Masuda}}, \bibinfo {author} {\bibfnamefont {M.~A.}\ \bibnamefont {Porter}},\
  and\ \bibinfo {author} {\bibfnamefont {R.}~\bibnamefont {Lambiotte}},\
  }\bibfield  {title} {\bibinfo {title} {Random walks and diffusion on
  networks},\ }\href {https://doi.org/10.1016/j.physrep.2017.07.007} {\bibfield
   {journal} {\bibinfo  {journal} {Phys. Rep.}\ }\textbf {\bibinfo {volume}
  {716}},\ \bibinfo {pages} {1} (\bibinfo {year} {2017})}\BibitemShut {NoStop}%
\bibitem [{\citenamefont {Bianconi}\ and\ \citenamefont
  {Dorogovstev}(2020)}]{Ginestra2020}%
  \BibitemOpen
  \bibfield  {author} {\bibinfo {author} {\bibfnamefont {G.}~\bibnamefont
  {Bianconi}}\ and\ \bibinfo {author} {\bibfnamefont {S.~N.}\ \bibnamefont
  {Dorogovstev}},\ }\bibfield  {title} {\bibinfo {title} {The spectral
  dimension of simplicial complexes: a renormalization group theory},\ }\href
  {https://doi.org/10.1088/1742-5468/ab5d0e} {\bibfield  {journal} {\bibinfo
  {journal} {J. Stat. Mech.: Theory Exp.}\ }\textbf {\bibinfo {volume}
  {2020}}\bibinfo  {number} { (1)},\ \bibinfo {pages} {014005}}\BibitemShut
  {NoStop}%
\bibitem [{\citenamefont {De~Domenico}\ and\ \citenamefont
  {Biamonte}(2016)}]{Domenico2016}%
  \BibitemOpen
\bibfield  {number} {  }\bibfield  {author} {\bibinfo {author} {\bibfnamefont
  {M.}~\bibnamefont {De~Domenico}}\ and\ \bibinfo {author} {\bibfnamefont
  {J.}~\bibnamefont {Biamonte}},\ }\bibfield  {title} {\bibinfo {title}
  {Spectral entropies as information-theoretic tools for complex network
  comparison},\ }\href {https://doi.org/10.1103/PhysRevX.6.041062} {\bibfield
  {journal} {\bibinfo  {journal} {Phys. Rev. X}\ }\textbf {\bibinfo {volume}
  {6}},\ \bibinfo {pages} {041062} (\bibinfo {year} {2016})}\BibitemShut
  {NoStop}%
\bibitem [{\citenamefont {Villegas}\ \emph {et~al.}(2022)\citenamefont
  {Villegas}, \citenamefont {Gabrielli}, \citenamefont {Santucci},
  \citenamefont {Caldarelli},\ and\ \citenamefont {Gili}}]{Villegas2022}%
  \BibitemOpen
  \bibfield  {author} {\bibinfo {author} {\bibfnamefont {P.}~\bibnamefont
  {Villegas}}, \bibinfo {author} {\bibfnamefont {A.}~\bibnamefont {Gabrielli}},
  \bibinfo {author} {\bibfnamefont {F.}~\bibnamefont {Santucci}}, \bibinfo
  {author} {\bibfnamefont {G.}~\bibnamefont {Caldarelli}},\ and\ \bibinfo
  {author} {\bibfnamefont {T.}~\bibnamefont {Gili}},\ }\bibfield  {title}
  {\bibinfo {title} {Path-integral approach to information processing in
  complex networks: information core emerges from entropic transitions},\
  }\href {https://arxiv.org/abs/2202.06669} {\bibfield  {journal} {\bibinfo
  {journal} {arXiv preprint arXiv:2202.06669}\ } (\bibinfo {year}
  {2022})}\BibitemShut {NoStop}%
\bibitem [{\citenamefont {Migdal}(1976)}]{migdal1976}%
  \BibitemOpen
  \bibfield  {author} {\bibinfo {author} {\bibfnamefont {A.~A.}\ \bibnamefont
  {Migdal}},\ }\bibfield  {title} {\bibinfo {title} {Phase transitions in gauge
  and spin-lattice systems},\ }\href@noop {} {\bibfield  {journal} {\bibinfo
  {journal} {J. Exp. Theor. Phys.}\ }\textbf {\bibinfo {volume} {42}},\
  \bibinfo {pages} {743} (\bibinfo {year} {1976})}\BibitemShut {NoStop}%
\bibitem [{\citenamefont {Newman}(2010)}]{Newman}%
  \BibitemOpen
  \bibfield  {author} {\bibinfo {author} {\bibfnamefont {M.~E.~J.}\
  \bibnamefont {Newman}},\ }\href
  {https://doi.org/10.1093/acprof:oso/9780199206650.001.0001} {\emph {\bibinfo
  {title} {Networks: an introduction}}}\ (\bibinfo  {publisher} {Oxford
  University Press},\ \bibinfo {address} {Oxford; New York},\ \bibinfo {year}
  {2010})\BibitemShut {NoStop}%
\bibitem [{\citenamefont {Feynman}\ \emph {et~al.}(2010)\citenamefont
  {Feynman}, \citenamefont {Hibbs},\ and\ \citenamefont {Styer}}]{Feynman}%
  \BibitemOpen
  \bibfield  {author} {\bibinfo {author} {\bibfnamefont {R.~P.}\ \bibnamefont
  {Feynman}}, \bibinfo {author} {\bibfnamefont {A.~R.}\ \bibnamefont {Hibbs}},\
  and\ \bibinfo {author} {\bibfnamefont {D.~F.}\ \bibnamefont {Styer}},\
  }\href@noop {} {\emph {\bibinfo {title} {Quantum mechanics and path
  integrals}}}\ (\bibinfo  {publisher} {Courier Corporation},\ \bibinfo
  {address} {Chelmsford},\ \bibinfo {year} {2010})\BibitemShut {NoStop}%
\bibitem [{\citenamefont {Moretti}\ and\ \citenamefont
  {Zaiser}(2019)}]{moretti2019}%
  \BibitemOpen
  \bibfield  {author} {\bibinfo {author} {\bibfnamefont {P.}~\bibnamefont
  {Moretti}}\ and\ \bibinfo {author} {\bibfnamefont {M.}~\bibnamefont
  {Zaiser}},\ }\bibfield  {title} {\bibinfo {title} {Network analysis predicts
  failure of materials and structures},\ }\href
  {https://doi.org/10.1073/pnas.1911715116} {\bibfield  {journal} {\bibinfo
  {journal} {Proc. Natl. Acad. Sci. U.S.A}\ }\textbf {\bibinfo {volume}
  {116}},\ \bibinfo {pages} {16666} (\bibinfo {year} {2019})}\BibitemShut
  {NoStop}%
\bibitem [{\citenamefont {Ghavasieh}\ \emph {et~al.}(2020)\citenamefont
  {Ghavasieh}, \citenamefont {Nicolini},\ and\ \citenamefont
  {De~Domenico}}]{ghavasieh2020}%
  \BibitemOpen
  \bibfield  {author} {\bibinfo {author} {\bibfnamefont {A.}~\bibnamefont
  {Ghavasieh}}, \bibinfo {author} {\bibfnamefont {C.}~\bibnamefont
  {Nicolini}},\ and\ \bibinfo {author} {\bibfnamefont {M.}~\bibnamefont
  {De~Domenico}},\ }\bibfield  {title} {\bibinfo {title} {Statistical physics
  of complex information dynamics},\ }\href
  {https://doi.org/10.1103/PhysRevE.102.052304} {\bibfield  {journal} {\bibinfo
   {journal} {Phys. Rev. E}\ }\textbf {\bibinfo {volume} {102}},\ \bibinfo
  {pages} {052304} (\bibinfo {year} {2020})}\BibitemShut {NoStop}%
\bibitem [{\citenamefont {Pathria}\ and\ \citenamefont
  {Beale}(2011)}]{Pathria}%
  \BibitemOpen
  \bibfield  {author} {\bibinfo {author} {\bibfnamefont {R.~K.}\ \bibnamefont
  {Pathria}}\ and\ \bibinfo {author} {\bibfnamefont {P.~D.}\ \bibnamefont
  {Beale}},\ }\href@noop {} {\emph {\bibinfo {title} {Statistical mechanics}}}\
  (\bibinfo  {publisher} {Elsevier/Academic Press},\ \bibinfo {address}
  {Amsterdam},\ \bibinfo {year} {2011})\BibitemShut {NoStop}%
\bibitem [{\citenamefont {Greiner}\ \emph {et~al.}(2012)\citenamefont
  {Greiner}, \citenamefont {Neise},\ and\ \citenamefont
  {St{\"o}cker}}]{Greiner}%
  \BibitemOpen
  \bibfield  {author} {\bibinfo {author} {\bibfnamefont {W.}~\bibnamefont
  {Greiner}}, \bibinfo {author} {\bibfnamefont {L.}~\bibnamefont {Neise}},\
  and\ \bibinfo {author} {\bibfnamefont {H.}~\bibnamefont {St{\"o}cker}},\
  }\href {https://doi.org/10.1007/978-1-4612-0827-3} {\emph {\bibinfo {title}
  {Thermodynamics and statistical mechanics}}}\ (\bibinfo  {publisher}
  {Springer},\ \bibinfo {address} {New York},\ \bibinfo {year}
  {2012})\BibitemShut {NoStop}%
\bibitem [{\citenamefont {Graham}(1977)}]{Graham1977}%
  \BibitemOpen
  \bibfield  {author} {\bibinfo {author} {\bibfnamefont {R.}~\bibnamefont
  {Graham}},\ }\bibfield  {title} {\bibinfo {title} {Path integral formulation
  of general diffusion processes},\ }\href {https://doi.org/10.1007/BF01312935}
  {\bibfield  {journal} {\bibinfo  {journal} {Z. Phys., B Condens. matter}\
  }\textbf {\bibinfo {volume} {26}},\ \bibinfo {pages} {281} (\bibinfo {year}
  {1977})}\BibitemShut {NoStop}%
\bibitem [{\citenamefont {Christensen}\ and\ \citenamefont
  {Moloney}(2005)}]{Moloney}%
  \BibitemOpen
  \bibfield  {author} {\bibinfo {author} {\bibfnamefont {K.}~\bibnamefont
  {Christensen}}\ and\ \bibinfo {author} {\bibfnamefont {N.~R.}\ \bibnamefont
  {Moloney}},\ }\href@noop {} {\emph {\bibinfo {title} {Complexity and
  criticality}}},\ Vol.~\bibinfo {volume} {1}\ (\bibinfo  {publisher} {World
  Scientific Publishing Company},\ \bibinfo {year} {2005})\BibitemShut
  {NoStop}%
\bibitem [{SI()}]{SI}%
  \BibitemOpen
  \href@noop {} {}\bibinfo {note} {See Supplementary Information at [] for
  further network analyses and examples on the coarse graining
  process.}\BibitemShut {Stop}%
\bibitem [{\citenamefont {Serafino}\ \emph {et~al.}(2021)\citenamefont
  {Serafino}, \citenamefont {Cimini}, \citenamefont {Maritan}, \citenamefont
  {Rinaldo}, \citenamefont {Suweis}, \citenamefont {Banavar},\ and\
  \citenamefont {Caldarelli}}]{Serafino2021}%
  \BibitemOpen
  \bibfield  {author} {\bibinfo {author} {\bibfnamefont {M.}~\bibnamefont
  {Serafino}}, \bibinfo {author} {\bibfnamefont {G.}~\bibnamefont {Cimini}},
  \bibinfo {author} {\bibfnamefont {A.}~\bibnamefont {Maritan}}, \bibinfo
  {author} {\bibfnamefont {A.}~\bibnamefont {Rinaldo}}, \bibinfo {author}
  {\bibfnamefont {S.}~\bibnamefont {Suweis}}, \bibinfo {author} {\bibfnamefont
  {J.~R.}\ \bibnamefont {Banavar}},\ and\ \bibinfo {author} {\bibfnamefont
  {G.}~\bibnamefont {Caldarelli}},\ }\bibfield  {title} {\bibinfo {title} {True
  scale-free networks hidden by finite size effects},\ }\bibfield  {journal}
  {\bibinfo  {journal} {Proc. Natl. Acad. Sci. U.S.A}\ }\textbf {\bibinfo
  {volume} {118}},\ \href {https://doi.org/10.1073/pnas.2013825118}
  {10.1073/pnas.2013825118} (\bibinfo {year} {2021})\BibitemShut {NoStop}%
\bibitem [{\citenamefont {Das}\ and\ \citenamefont {Yu}(2012)}]{das2012}%
  \BibitemOpen
  \bibfield  {author} {\bibinfo {author} {\bibfnamefont {J.}~\bibnamefont
  {Das}}\ and\ \bibinfo {author} {\bibfnamefont {H.}~\bibnamefont {Yu}},\
  }\bibfield  {title} {\bibinfo {title} {Hint: High-quality protein
  interactomes and their applications in understanding human disease},\ }\href
  {https://doi.org/10.1186/1752-0509-6-92} {\bibfield  {journal} {\bibinfo
  {journal} {BMC Syst. Biol.}\ }\textbf {\bibinfo {volume} {6}},\ \bibinfo
  {pages} {1} (\bibinfo {year} {2012})}\BibitemShut {NoStop}%
\bibitem [{\citenamefont {Huss}\ and\ \citenamefont {Holme}(2007)}]{Dros}%
  \BibitemOpen
  \bibfield  {author} {\bibinfo {author} {\bibfnamefont {M.}~\bibnamefont
  {Huss}}\ and\ \bibinfo {author} {\bibfnamefont {P.}~\bibnamefont {Holme}},\
  }\bibfield  {title} {\bibinfo {title} {Currency and commodity metabolites:
  their identification and relation to the modularity of metabolic networks},\
  }\href {https://doi.org/https://doi.org/10.1049/iet-syb:20060077} {\bibfield
  {journal} {\bibinfo  {journal} {IET Syst. Biol.}\ }\textbf {\bibinfo {volume}
  {1}},\ \bibinfo {pages} {280} (\bibinfo {year} {2007})}\BibitemShut {NoStop}%
\bibitem [{\citenamefont {Harris}\ \emph {et~al.}(1975)\citenamefont {Harris},
  \citenamefont {Lubensky}, \citenamefont {Holcomb},\ and\ \citenamefont
  {Dasgupta}}]{Harris1975}%
  \BibitemOpen
  \bibfield  {author} {\bibinfo {author} {\bibfnamefont {A.~B.}\ \bibnamefont
  {Harris}}, \bibinfo {author} {\bibfnamefont {T.~C.}\ \bibnamefont
  {Lubensky}}, \bibinfo {author} {\bibfnamefont {W.~K.}\ \bibnamefont
  {Holcomb}},\ and\ \bibinfo {author} {\bibfnamefont {C.}~\bibnamefont
  {Dasgupta}},\ }\bibfield  {title} {\bibinfo {title} {Renormalization-group
  approach to percolation problems},\ }\href
  {https://doi.org/10.1103/PhysRevLett.35.327} {\bibfield  {journal} {\bibinfo
  {journal} {Phys. Rev. Lett.}\ }\textbf {\bibinfo {volume} {35}},\ \bibinfo
  {pages} {327} (\bibinfo {year} {1975})}\BibitemShut {NoStop}%
\bibitem [{\citenamefont {De~Gennes}\ and\ \citenamefont
  {Gennes}(1979)}]{DeGennes1979}%
  \BibitemOpen
  \bibfield  {author} {\bibinfo {author} {\bibfnamefont {P.-G.}\ \bibnamefont
  {De~Gennes}}\ and\ \bibinfo {author} {\bibfnamefont {P.-G.}\ \bibnamefont
  {Gennes}},\ }\href@noop {} {\emph {\bibinfo {title} {Scaling concepts in
  polymer physics}}}\ (\bibinfo  {publisher} {Cornell university press},\
  \bibinfo {address} {New York},\ \bibinfo {year} {1979})\BibitemShut {NoStop}%
\bibitem [{\citenamefont {Villa~Mart{\'\i}n}\ \emph {et~al.}(2015)\citenamefont
  {Villa~Mart{\'\i}n}, \citenamefont {Bonachela}, \citenamefont {Levin},\ and\
  \citenamefont {Mu{\~n}oz}}]{PVilla}%
  \BibitemOpen
  \bibfield  {author} {\bibinfo {author} {\bibfnamefont {P.}~\bibnamefont
  {Villa~Mart{\'\i}n}}, \bibinfo {author} {\bibfnamefont {J.~A.}\ \bibnamefont
  {Bonachela}}, \bibinfo {author} {\bibfnamefont {S.~A.}\ \bibnamefont
  {Levin}},\ and\ \bibinfo {author} {\bibfnamefont {M.~A.}\ \bibnamefont
  {Mu{\~n}oz}},\ }\bibfield  {title} {\bibinfo {title} {Eluding catastrophic
  shifts},\ }\href {https://doi.org/10.1073/pnas.1414708112} {\bibfield
  {journal} {\bibinfo  {journal} {Proc. Natl. Acad. Sci. U.S.A}\ }\textbf
  {\bibinfo {volume} {112}},\ \bibinfo {pages} {E1828} (\bibinfo {year}
  {2015})}\BibitemShut {NoStop}%
\bibitem [{\citenamefont {Meshulam}\ \emph {et~al.}(2019)\citenamefont
  {Meshulam}, \citenamefont {Gauthier}, \citenamefont {Brody}, \citenamefont
  {Tank},\ and\ \citenamefont {Bialek}}]{meshulam2019}%
  \BibitemOpen
  \bibfield  {author} {\bibinfo {author} {\bibfnamefont {L.}~\bibnamefont
  {Meshulam}}, \bibinfo {author} {\bibfnamefont {J.~L.}\ \bibnamefont
  {Gauthier}}, \bibinfo {author} {\bibfnamefont {C.~D.}\ \bibnamefont {Brody}},
  \bibinfo {author} {\bibfnamefont {D.~W.}\ \bibnamefont {Tank}},\ and\
  \bibinfo {author} {\bibfnamefont {W.}~\bibnamefont {Bialek}},\ }\bibfield
  {title} {\bibinfo {title} {Coarse graining, fixed points, and scaling in a
  large population of neurons},\ }\href
  {https://doi.org/10.1103/PhysRevLett.123.178103} {\bibfield  {journal}
  {\bibinfo  {journal} {Phys. Rev. Lett.}\ }\textbf {\bibinfo {volume} {123}},\
  \bibinfo {pages} {178103} (\bibinfo {year} {2019})}\BibitemShut {NoStop}%
\bibitem [{\citenamefont {Cavagna}\ \emph {et~al.}(2019)\citenamefont
  {Cavagna}, \citenamefont {Di~Carlo}, \citenamefont {Giardina}, \citenamefont
  {Grandinetti}, \citenamefont {Grigera},\ and\ \citenamefont
  {Pisegna}}]{Cavagna2019}%
  \BibitemOpen
  \bibfield  {author} {\bibinfo {author} {\bibfnamefont {A.}~\bibnamefont
  {Cavagna}}, \bibinfo {author} {\bibfnamefont {L.}~\bibnamefont {Di~Carlo}},
  \bibinfo {author} {\bibfnamefont {I.}~\bibnamefont {Giardina}}, \bibinfo
  {author} {\bibfnamefont {L.}~\bibnamefont {Grandinetti}}, \bibinfo {author}
  {\bibfnamefont {T.~S.}\ \bibnamefont {Grigera}},\ and\ \bibinfo {author}
  {\bibfnamefont {G.}~\bibnamefont {Pisegna}},\ }\bibfield  {title} {\bibinfo
  {title} {Dynamical renormalization group approach to the collective behavior
  of swarms},\ }\href {https://doi.org/10.1103/PhysRevLett.123.268001}
  {\bibfield  {journal} {\bibinfo  {journal} {Phys. Rev. Lett.}\ }\textbf
  {\bibinfo {volume} {123}},\ \bibinfo {pages} {268001} (\bibinfo {year}
  {2019})}\BibitemShut {NoStop}%
\bibitem [{\citenamefont {Cohen}\ and\ \citenamefont
  {Havlin}(2003)}]{Cohen2003}%
  \BibitemOpen
  \bibfield  {author} {\bibinfo {author} {\bibfnamefont {R.}~\bibnamefont
  {Cohen}}\ and\ \bibinfo {author} {\bibfnamefont {S.}~\bibnamefont {Havlin}},\
  }\bibfield  {title} {\bibinfo {title} {Scale-free networks are ultrasmall},\
  }\href {https://doi.org/10.1103/PhysRevLett.90.058701} {\bibfield  {journal}
  {\bibinfo  {journal} {Phys. Rev. Lett.}\ }\textbf {\bibinfo {volume} {90}},\
  \bibinfo {pages} {058701} (\bibinfo {year} {2003})}\BibitemShut {NoStop}%
\bibitem [{\citenamefont {Marro}\ and\ \citenamefont {Dickman}(1999)}]{Marro}%
  \BibitemOpen
  \bibfield  {author} {\bibinfo {author} {\bibfnamefont {J.}~\bibnamefont
  {Marro}}\ and\ \bibinfo {author} {\bibfnamefont {R.}~\bibnamefont
  {Dickman}},\ }\href@noop {} {\emph {\bibinfo {title} {{Nonequilibrium Phase
  Transition in Lattice Models}}}}\ (\bibinfo  {publisher} {Cambridge
  University Press},\ \bibinfo {address} {Cambridge},\ \bibinfo {year}
  {1999})\BibitemShut {NoStop}%
\bibitem [{\citenamefont {Gardiner}(2009)}]{Gardiner}%
  \BibitemOpen
  \bibfield  {author} {\bibinfo {author} {\bibfnamefont {C.}~\bibnamefont
  {Gardiner}},\ }\href {https://doi.org/10.1002/bbpc.19850890629} {\emph
  {\bibinfo {title} {Stochastic Methods: A Handbook for the Natural and Social
  Sciences}}},\ Vol.~\bibinfo {volume} {4}\ (\bibinfo  {publisher} {Springer},\
  \bibinfo {address} {Berlin},\ \bibinfo {year} {2009})\BibitemShut {NoStop}%
\end{thebibliography}

\clearpage


\section*{Supplementary material}

# Supplementary Information: Laplacian Renormalization Group for heterogeneous networks

Pablo Villegas,[1] Tommaso Gili,[2] Guido Caldarelli,[3,4,5,6,*] and Andrea Gabrielli[1,7,5]

[1]*'Enrico Fermi' Research Center (CREF), Via Panisperna 89A, 00184 - Rome, Italy*
[2]*IMT Institute for Advanced Studies, Piazza San Ponziano 6, 55100 Lucca, Italy.*
[3]*Department of Molecular Sciences and Nanosystems,*
*Ca' Foscari University of Venice, 30172 Venice, Italy*
[4]*European Centre for Living Technology, 30124 Venice, Italy*
[5]*Institute for Complex Systems, Consiglio Nazionale delle Ricerche, UoS Sapienza, 00185 Rome, Italy*
[6]*London Institute for Mathematical Sciences, W1K2XF London, United Kingdom*
[7]*Dipartimento di Ingegneria, Università Roma Tre, 00146, Rome, Italy*

## CONTENTS

---

* Corresponding author: guido.caldarelli@unive.it

## 1. REAL-SPACE RENORMALIZATION GROUP

The development of the renormalization group is a significant milestone that allowed for dealing with some of the most challenging problems of physics, ranging from relativistic quantum field theory to critical phenomena or the Kondo effect. Moreover, it allowed solving two critical problems in physics, namely, the task of solving systems with many degrees of freedom and the description of how the cooperative features of collective behavior arise [1]. Here we briefly describe the key RG concepts, making the analogy with RG diffusion on complex networks easier.

Let us consider a Hamiltonian $\mathcal{H}(\sigma(x))$, where $\sigma(x)$ represents the field of local mean spin and thus is a continuum variable. The main idea of the RG is to find an *effective Hamiltonian* at scale $\lambda$, $\mathcal{H}(\sigma) \mapsto \mathcal{H}_\lambda(\sigma)$.

One of the ideas to develop the RG process was proposed in 1966 by Kadanoff to deal with the real-space RG problem. Let us illustrate the procedure for the specific case of the Ising model. The three-step procedure (illustrated in Figure 1) to perform RG in real-space is:

1. Group the lattice points into groups of $b$ sites.

2. Replace each block with a single one of size $b$, where $\mathcal{R}_g(\sigma)$ represents some renormalization group transformation as, e.g., the majority rule.

3. Rescale all scales by a factor $b$ to return to the original lattice spacing.

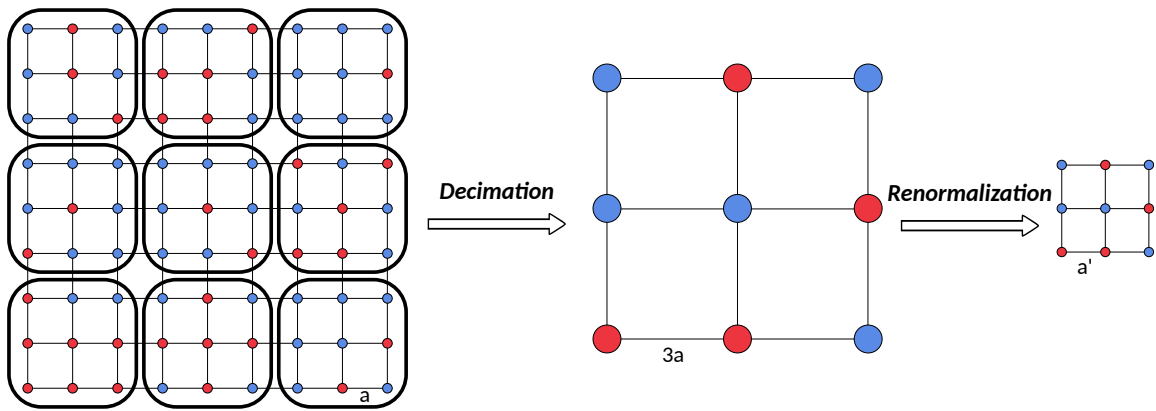

FIG. 1: Sketch the decimation process employing Kadanoff blocks on a square lattice of side $a$, where blue (red) points represent up (down) spins, respectively. The lattice is divided into several blocks with $b^2$ sites which are now coarse-grained, replacing it with a single-block following some rule $\mathcal{R}_g(\sigma)$, finally reducing all the system scales by a factor $b$. This scheme produces a reduced version of the original system.

In the paradigmatic case of the Ising model, the renormalization group operates at any value of the temperature, $T$. However, the Kadanoff procedure induces particular flows towards the trivial fixed points $T = 0$ for any spin configuration with $T < T_c$ and $T = \infty$ for any spin configuration with $T > T_c$. For $T = T_c$, there is no flow associated with RG transformations [2], and it is a repulsive fixed point of the flow along with the (tuning) control parameter, which reflects the self-similar properties of the system. Figure 2 shows the particular application of Kadanoff blocks to the Ising model for the three possible cases: $T < T_c$, $T = T_c$, and $T > T_c$.

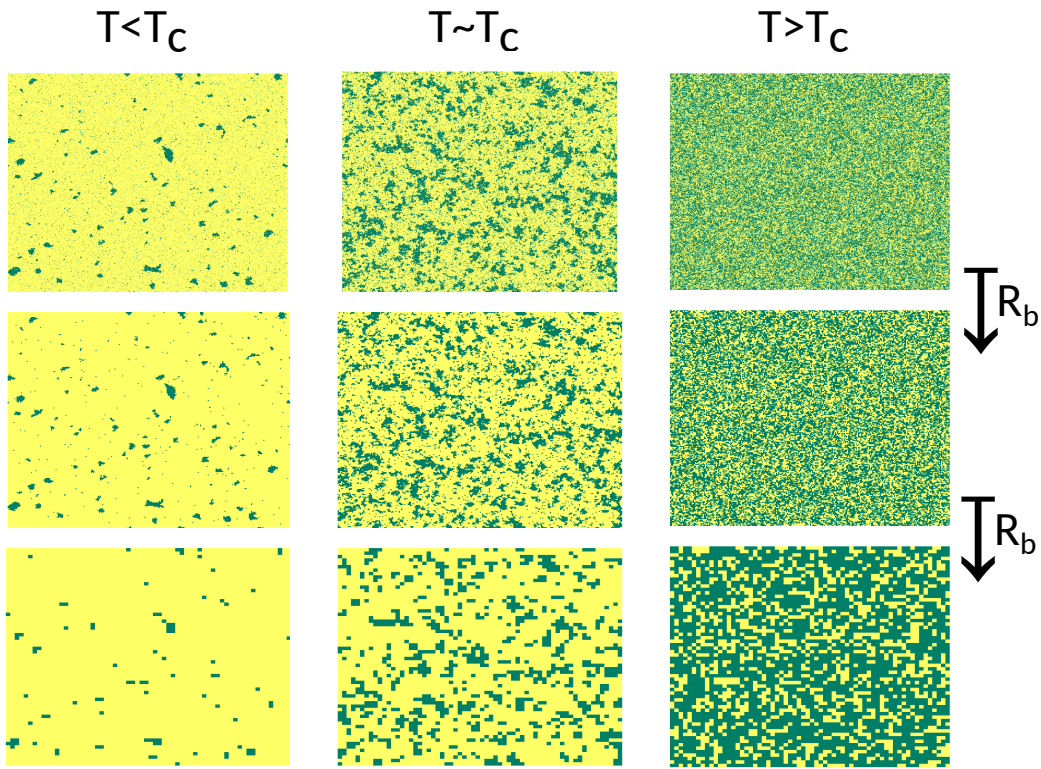

FIG. 2: (a) RG process for the Ising model simulated in a 2D lattice of $L = 1024$ (upper panels), and employing Kadanoff blocks of size $b = 3$. The three panels correspond to different temperatures $T < T_c$, $T \sim T_c$, and $T > T_c$, respectively. The RG process is performed from top to bottom scales of the figure, revealing the RG flow of the dynamical process.

Similarly, we expect to see RG flow when performing RG transformations on top of networks lacking scale-free properties. Hence, to gain analytical insight into the RG process employing Kadanoff supernodes, here we analyze the case of pertinent scale-dependent networks, i.e., those who present an isolate resolution single scale being responsible for the physical behavior of macroscopic quantities as, e.g., Erdös-Renyi networks or stochastic block models (SBM). In complete analogy with Kadanoff transformations, such networks are expected to follow the RG towards a trivial fixed point of the network structure, yet reflecting crucial mesoscales.

4

## 2. SCALE-DEPENDENT NETWORKS

### 2.1. Erdös-Renyi network

Figure 3(b) shows the diffusion of information over the network for simple Erdös-Rényi graphs. The graph's particular structure, lacking any sign of scale-invariance, is reflected in the single-peal of $C$, where information diffuses all over the network. Note that any Kadanoff transformation in the vicinity of that peak will automatically collapse the whole network into a single node, capturing the arguments described in the main text. It is possible to perform some RG steps before this point, constantly changing the network properties. Figure 3(a) shows a particular case for successive RG transformations with $\tau = 0.25$ (within the greyish area of Figure 3(b)), reflecting two types of effects: (i) Network properties continuously change, evidencing the flowing towards a trivial fixed point. (ii) ER networks converge rapidly to the trivial fixed point where the network acts as a single node.

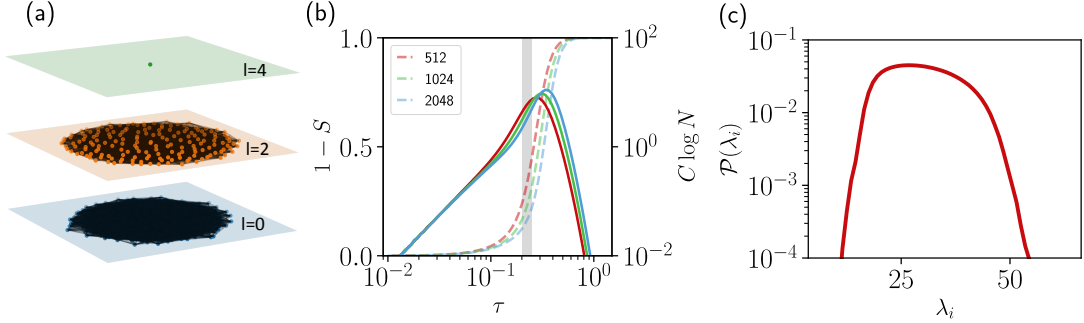

FIG. 3: (a) RG process for Erdös-Renyi networks of $N = 256$ employing $\tau = 0.25$. Each layer represents the RG step of order $l$. (b) Entropy parameter (dashed lines, $(1-S)$), and specific heat (solid lines, $C$), versus the temporal resolution parameter of the network, $\tau$ for Erdös-Renyi networks of different systems sizes (see legend) with mean connectivity $\langle k \rangle = 30$. (c) Probability distribution of Laplacian eigenvalues for an Erdös-Renyi network with $N = 4096$ and $\langle k \rangle = 30$.

In particular, Figure 4 illustrates the evolution of the averaged degree distribution over network realizations, $P(\kappa)$, for subsequent RG transformations and different resolution scales, $\tau$. Note that, as soon as any coarse grain is performed, the mean connectivity flows undoubtedly towards smaller values, according to the results presented in the main text, until it collapses in a trivial fixed point, as shown in Figure 3(a) and Figure 4(c).

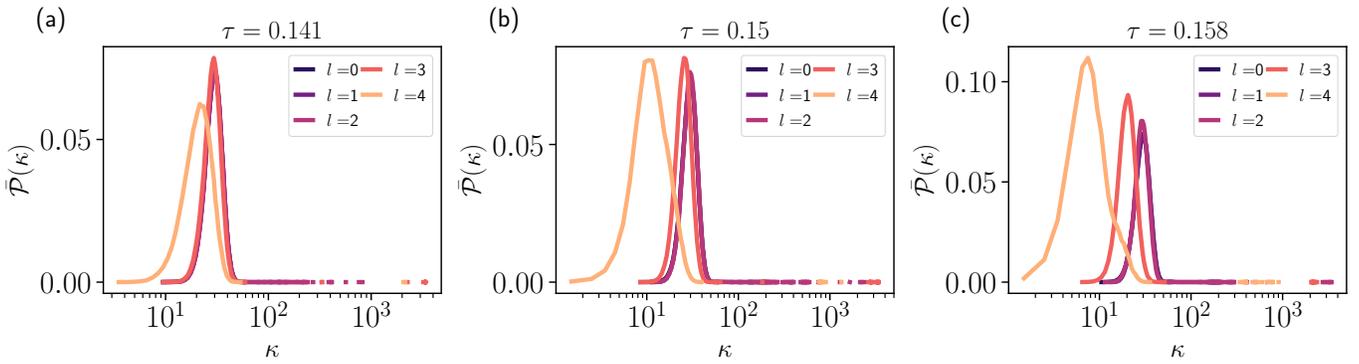

FIG. 4: Degree distribution after subsequent RG transformations for an Erdös-Rényi network with $N = 4096$ nodes and mean connectivity $\kappa = \langle 30 \rangle$ for different values of $\tau$ (see title): (a) $\tau = 0.141$, (b) $\tau = 0.15$, and (c) $\tau = 0.158$. Note that, as soon as we perform any blocking of nodes, the degree distribution, $P(\kappa)$, suddenly changes flowing through a trivial fixed point. All curves have been averaged over $10^2$ network realizations.

## 2.2. Stochastic-block model

Despite the network convergence through a trivial fixed point due to its intrinsic lacking of any scale-invariance, the study of RG flow in networks with characteristic scales can shed light on its characteristic scales and functional modules. We stress here the particular case of a stochastic block model, composed of four different modules and different intra and interconnectivity between them. As shown in Figure 5, a detailed analysis close to $\tau^*$ makes emerge the four modules at a mesoscopic scale at some point of the RG process. Finally, the network will always collapse at a trivial fixed point.

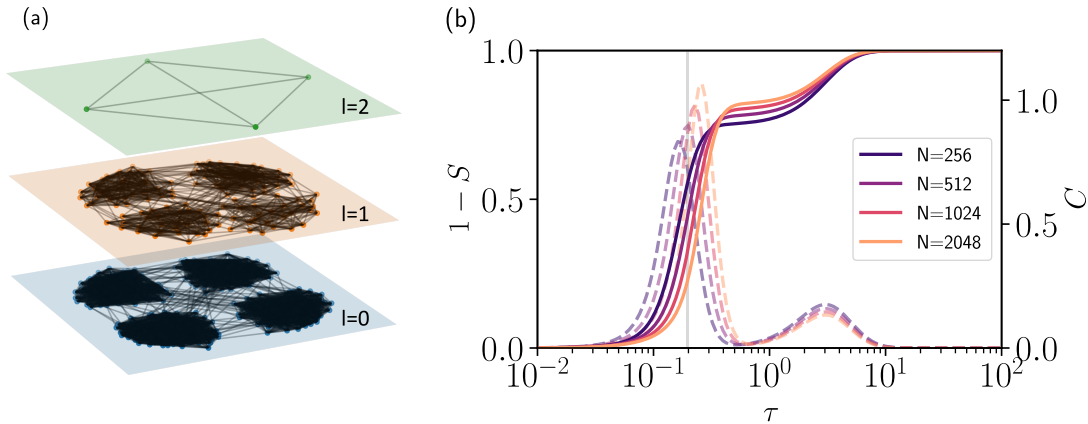

FIG. 5: (a) RG process for a SBM of size $N = 256$ employing $\tau = 0.2$. Each layer represents the RG step of order $l$. (b) Entropy parameter for the stochastic block model considering different network sizes (see legend) constituted by four equal interconnected modules $p = 128/N$ and interconnectivity probability $q = 1/N$. The two peaks in the derivative of the order parameter $C$ indicate the two critical network resolution scales.

## 2.3. Small-world networks

One of the main problems in analyzing RG procedures, or network reduction, is to deal with small-world or ultra small-world problems giving place to short path length between all nodes in the network structure, thus generating an infinite spectral dimension. Here we show the capability of our method to perform networks reduction even in the worst scenario where network scales are strongly interlinked.

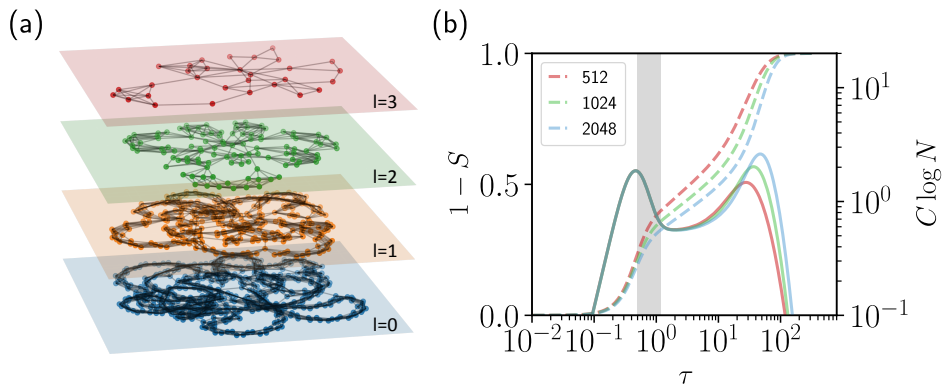

FIG. 6: (a) RG process for a SW of size $N = 512$ employing $\tau = 0.8$. Each layer represents the RG step of order $l$. (b) Entropy parameter for a SW network with rewiring probability $p = 0.02$ considering different network sizes (see legend). The two peaks in the specific heat, $C$, indicate the two critical network resolution scales: the underlying lattice and the long-range connections which allow for information to be distributed and processed across the entire network.

Figure 6(a) shows the RG procedure for a small-world network [3] setting a certain probability of long-range connections which allow for information to be distributed and processed across the entire network. Note the particular shape of the entropy parameter at Figure 6(b). Regardless of the discussions regarding the scale-invariant nature of small-world networks, let us pinpoint here a critical issue. The way these networks are built strongly constrain the large-time specific heat peak. Note that its particular position reduces in time due to the effects of the long-range connections, even if the network's local scales (a regular lattice in this case) remain unaltered. The same effect is expected to occur in more complex structures, where local system' properties will determine the system's dynamics while small-world effects facilitate the information to be distributed and processed across the entire network in shorter times. In summary, our RG scheme allows us individuating different mesoscales and communities even in the presence of small-world substantial effects.

## 3. SCALE-INVARIANT NETWORKS

### 3.1. Random trees

We also analyze the simple case of simple trees and random trees, presenting complementary results to those of the main text. In particular, Figure 7 shows the RG diffusion scheme when applied over these two network structures. As it can be seen, both structures remain scale-invariant after subsequent RG transformations.

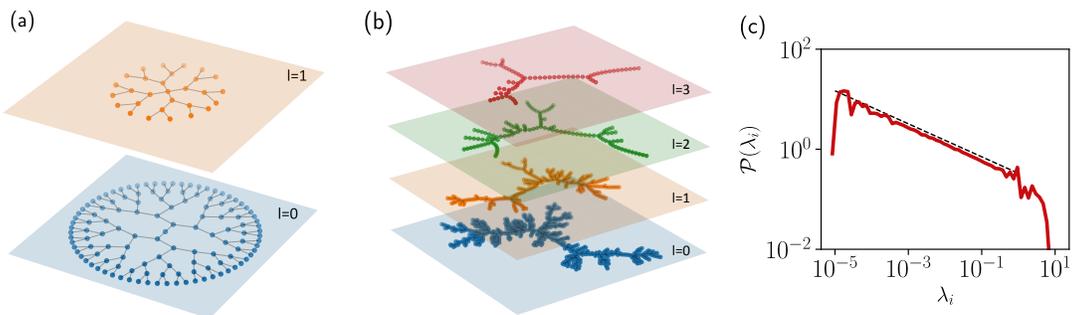

FIG. 7: (a) RG process for a regular tree with branching factor $r = 2$, and $h = 6$ hierarchical levels. (b) Four-step RG transformation for a random tree with $N = 1024$ nodes using $\tau = 3.5$. (c) Probability distribution of Laplacian eigenvalues for random tree with $N = 4096$. Dashed line is a guide to the eye with power law exponent $\gamma = -1/3$.

Figure 8 illustrates the evolution of the averaged degree distribution over network realizations, $P(\kappa)$, for subsequent RG transformations and different resolution scales, $\tau$. As we expected, the connectivity distribution remains invariant, according to the results presented in the main text.

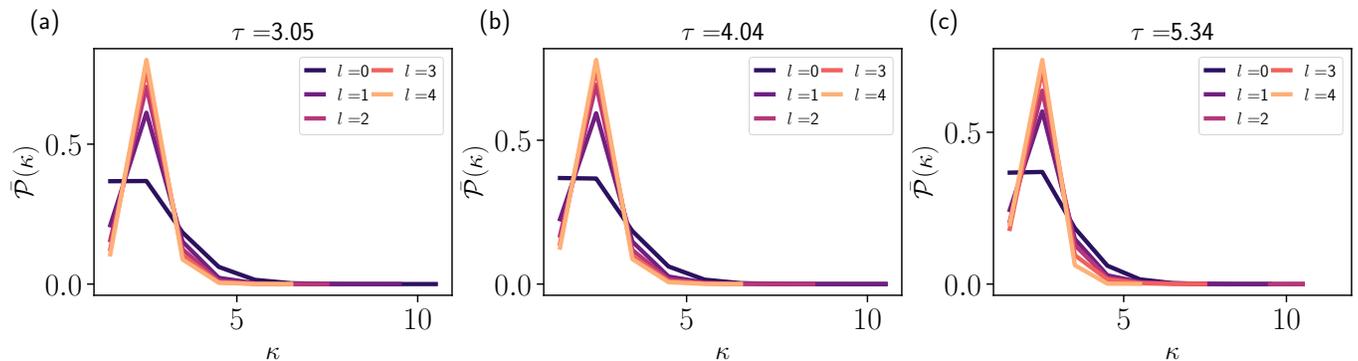

FIG. 8: Degree distribution after subsequent RG transformations for random trees with $N = 4096$ nodes for different values of $\tau$ (see title): (a) $\tau = 3.05$, (b) $\tau = 4.04$, and (c) $\tau = 5.34$. The degree distribution, $P(\kappa)$, remains invariant under succesive RG transformations. All curves have been averaged over $10^2$ network realizations.

## 3.2. 2D Lattices

For the sake of consistency and to avoid spurious results, we study here the simplest trivial scale-invariant structure: a regular two-dimensional lattice. As shown in Figure 9(b), a small $\tau^*$ peak reflects the characteristic resolution scale of the system, allowing us to perform the RG procedure over the lattice reducing it to smaller lattices for several RG steps (see Figure 9(a)). Note that RG transformations are consistent over all the greyish areas of Figure 9(b), after which Kadanoff blocks become too large and finite-size effects can alter renormalization process.

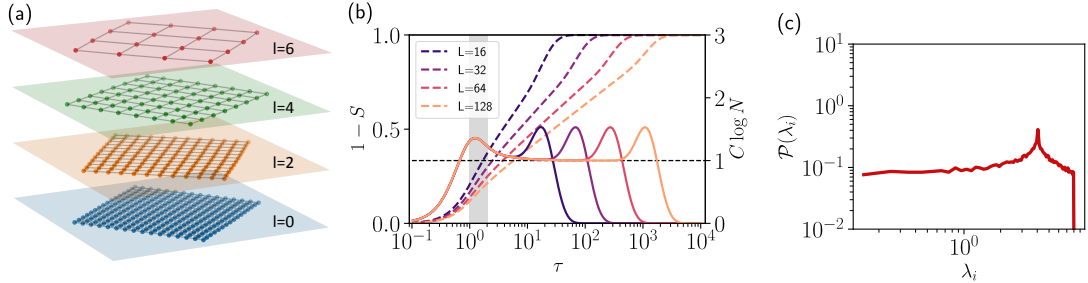

FIG. 9: (a) RG process for a 2D lattice of size $N = 16^2$ employing $\tau = 1.5$. Each layer represents the RG step of order $l$. (b) Entropy parameter for the 2D lattices considering different lattice sizes ($N = L^2$, see legend). The first peak allows us to identify the critical network scale to perform the RG reduction. (c) Probability distribution of Laplacian eigenvalues for a 2D lattice of size $N = 128^2$. Note that the spectrum has a characteristic scale of $\lambda = 4$ while presents a flat region that influence the particular value $C = 1$.

## 4. SCALE-FREE NETWORKS

### 4.1. Barabasi-Albert networks

We present here analyses of different Barabasi-Albert networks modifying the number of connecting nodes at each time-step in the network building procedure ($m$), which it is known to critically constrain the spectral dimension of the network. Figure 10(a) and (b) shows the evolution of the entropy parameter as a function of the resolution scale, $\tau$, for different system sizes. Pay close attention to the fact that information flows much faster all across the network due to small-world effects, and $\tau^*$, the peak at short-time scales, disappears. Despite this, it is yet possible to robustly perform network RG transformations even if the networks do not present scale-invariant properties anymore.

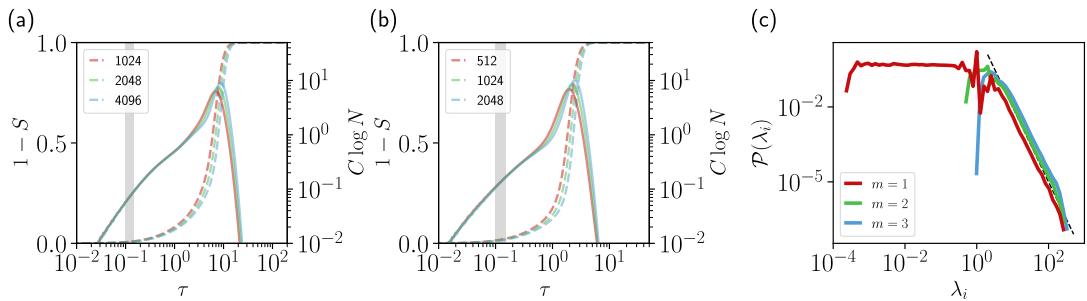

FIG. 10: (a) Entropy parameter (dashed lines, $(1 - S)$), and specific heat (solid lines, $C$), versus the temporal resolution parameter of the network, $\tau$ for BA networks of different systems sizes (see legend) with: (a) $m = 2$, and (b) $m = 4$. (c) Probability distribution of Laplacian eigenvalues for a BA with $N = 4096$ and different $m$ values. Note that only for $m = 1$ the BA network exhibits a flat region leading to a constant value of the network' specific heat. All curves have been averaged over $10^2$ network realizations.

As shown in Figure 11 (a) and (b) the degree distribution, $P(\kappa)$ remains invariant when we perform subsequent RG transformations over the network, also conserving the mean-connectivity and the different intrinsic network properties.

7

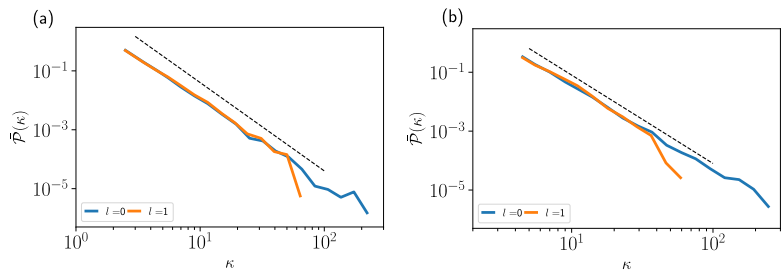

FIG. 11: Degree distribution after subsequent RG transformations for Barabasi-Albert networks with $N = 4096$ nodes and $\tau = 0.1$ for different values of $m$: (a) $m = 2$, (b) $m = 4$. The degree distribution, $P(\kappa)$, remains invariant under RG transformations. All curves have been averaged over $10^2$ network realizations.

## 5. REAL-WORLD NETWORKS

For the sake of completeness we present here a full analysis of different real scale-free metabolic networks of different organisms from a collection of real network data from the ICON at https://icon.colorado.edu/ as well as the KONECT at https://west.uni-koblenz.de/konect. Figure 12 shows the evolution of the entropy parameter as a function of the resolution scale, $\tau$, evidencing the complex mesoscopic structure of these networks. We also show the Laplacian spectrum, which evidence that only the $M.Musculus$ and $E.Coli$ network present clean scale-invariant properties, i.e. they present constant or power-law regimes in the eigenvalues spectrum. In any case, we can produce replicas for all the networks, with consistent degree distribution for the downscaled networks.

9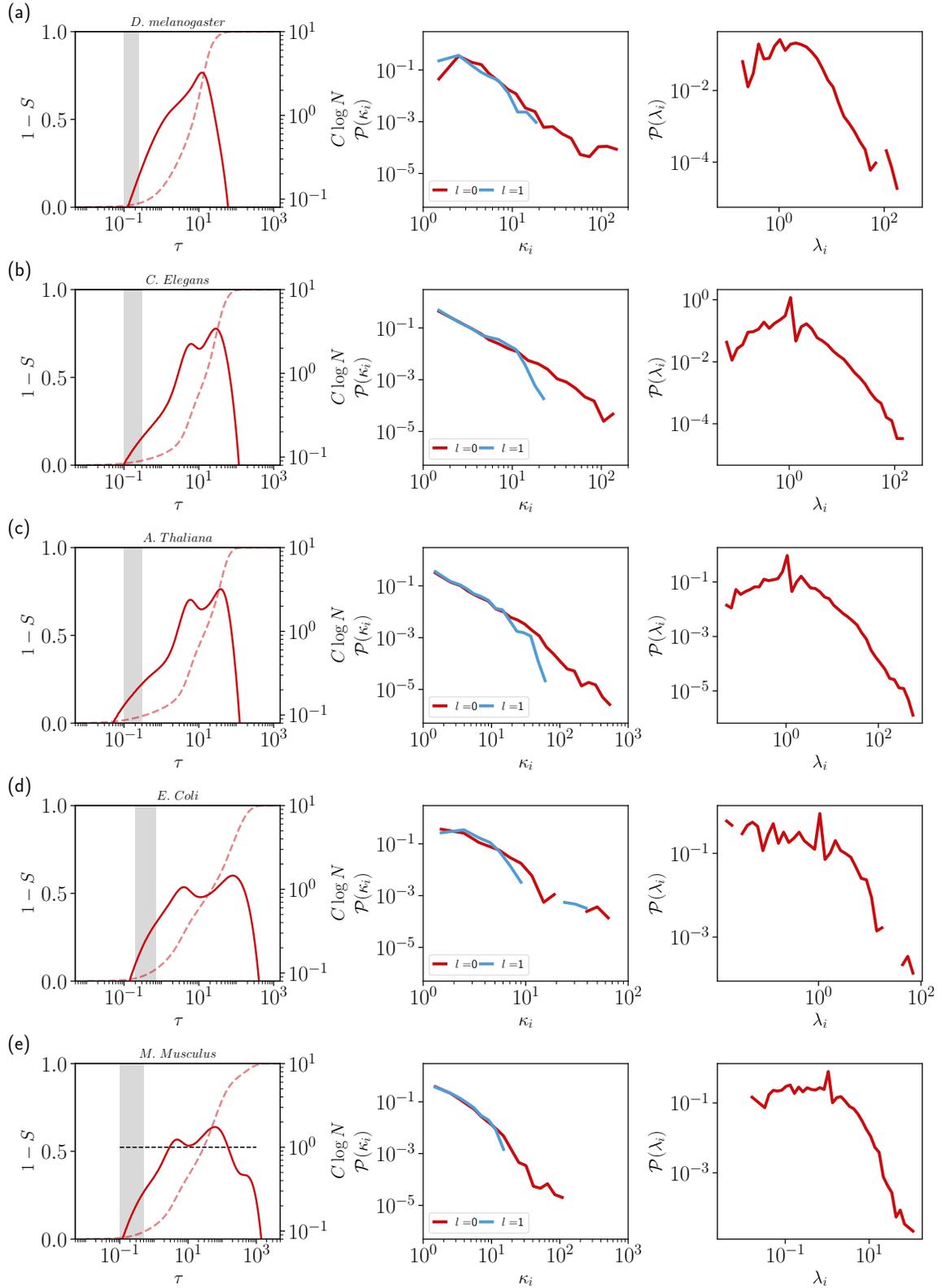

FIG. 12: **(Left column)** Entropy parameter (dashed lines, $(1-S)$), and specific heat (solid lines, $C$), versus the temporal resolution parameter of the network, $\tau$ for different species (see title). **(Central column)** RG transformation for the different considered species using: (a) $\tau = 0.2$ (b) $\tau = 0.2$ (c) $\tau = 0.1$ (d) $\tau = 0.5$ (e) $\tau = 0.3$ **(Right column)** Spectral probability distribution of the Laplacian matrix for the different species.

# REFERENCES

[1] K. G. Wilson and J. Kogut, The renormalization group and the $\epsilon$ expansion, Phys. Rep. **12**, 75 (1974).
[2] K. Christensen and N. R. Moloney, *Complexity and criticality*, Vol. 1 (World Scientific Publishing Company, 2005).
[3] D. J. Watts and S. H. Strogatz, Collective dynamics of 'small-world' networks, Nature **393**, 440 (1998).


\end{document}